\documentclass[aps,amsmath,amssymb,prl,showpacs,floatfix]{revtex4}
\usepackage{bm,natbib}
\usepackage[dvips]{graphicx}

\begin{document}

\title{BCS Superconductivity in Quantum Critical Metals}

\author{Jian-Huang She, and Jan Zaanen}

\affiliation{Instituut-Lorentz for Theoretical Physics, Universiteit Leiden, P. O. Box 9506, 2300 R A Leiden, The Netherlands}

\begin{abstract}
We present a simple phenomenological scaling theory for the pairing instability of a quantum critical metal. It
can be viewed as a minimal generalization of the classical Bardeen-Cooper-Schrieffer theory of superconductivity for
normal Fermi-liquid metals. We assume that attractive interactions are induced in the fermion system by an external 'bosonic glue' 
that is strongly retarded. Resting on the small Migdal parameter, all the required information from the fermion system needed
to address the superconductivity enters through the
pairing susceptibility. Asserting that the normal state is a  strongly interacting quantum critical state of fermions, the form of this susceptibility
is governed by conformal invariance and one only has the scaling dimension of the pair operator as free parameter. Within this scaling
framework, conventional BCS  theory appears as the 'marginal' case but it is now easily generalized to the (ir)relevant scaling regimes. 
In the relevant regime an algebraic singularity takes over from the BCS logarithm with the obvious effect that the pairing instability becomes
stronger. However, it is more surprising that this effect is strongest for small couplings and small Migdal parameters, highlighting an unanticipated
important role of retardation. Using exact forms for the finite temperature pair susceptibility from 1+1D conformal field theory as models, we
study the transition temperatures, finding that the gap to transition temperature ratio's are generically large compared to the BCS case, showing however an
opposite trend as function of the coupling strength compared to conventional Migdal-Eliashberg theory.  We show that our scaling theory naturally produces
the  superconducting 'domes' 
surrounding the quantum critical points,  even when the coupling to the glue itself is not changing at all.  We argue that hidden relations will
exist between the location of the cross-over lines to the Fermi-liquids away from the quantum critical points , and the detailed form of the dome
when the glue strength is independent of the zero temperature control parameter. Finally, we discuss the behavior of the orbital limited
upper critical magnetic field as function of the zero temperature coupling constant. Compared to the variation of the transition temperature, the 
critical field might show a much stronger variation pending the value of the dynamical critical exponent.  

\end{abstract}

\date{\today \ [file: \jobname]}

 \maketitle

\section{Introduction}
The 'mystery superconductors' of current
interest share the property that their normal states are poorly understood 'non Fermi-liquids'.  Experiments reveal that these are
governed by a scale invariance of their quantum dynamics. The best documented examples are found in the heavy fermion (HF) 
systems\cite{Lonzarich98, Zaanen08, Stewart06,Lohneisen07,Coleman01,Coleman05,Gegenwart08,Paschen04,Si01}. As function
of pressure or magnetic field one can drive a magnetic phase transition to zero temperature. On both sides of this quantum critical point (QCP) one finds
Fermi-liquids  characterized by quasiparticle masses that tend to diverge at the QCP. At the QCP one finds a 'strange metal' revealing traits of
scale invariance, while at a 'low' temperature a transition follows
most often to a superconducting state with a maximum $T_c$ right at the QCP. It is widely believed that a similar 'fermionic quantum criticality' is 
governing the normal state in optimally doped cuprate high Tc superconductors. The best evidence is perhaps the 'Planckian' relaxation time observed in
transport experiments $\tau_{\hbar} \simeq \hbar / (k_B T)$\cite{Marel03,Hussey09} indicating that this normal state has no knowledge of the scale $E_F$ 
since in a Fermi-liquid $\tau = (E_F/k_B T) \tau_{\hbar}$. Very recently indications have been found that even the iron based superconductors
might be governed by quantum critical normal states associated with a magnetic and/or structural zero temperature transition, giving rise to 
a novel scaling behavior of the electronic specific heat\cite{Budko09,Zaanen09}.  

The idea that superconductivity can be caused by a quantum phase transition involving a 
bosonic order parameter has a long history, starting with the marginal Fermi-liquid ideas of Varma\cite{Varma89} in the context of cuprates of 
the late 1980's and the ideas of spin-fluctuation driven heavy fermion superconductivity  dating back to Lonzarich {\em et al.}\cite{Lonzarich98}. The bulk of the 
large theoretical literature\cite{Monthoux07,Chubukov93,Varma02,Nayak96, Sachdev09,Chubukov05,Chubukov07,Chubukov01,Chubukov012,Chubukov03,Chubukov0602,Chubukov06,Chubukov08,Shively06,Sachdev0902,Fisk98,Mazin97,Monthoux99,Fay80,Millis88,Millis98,Millis01,Bedell99,Bedell01,Allen75,Marsiglio86,Carbotte90,Scalapino86,Bulaevskii90,Kirkpatrick01,Lonzarich03,Metzner09,Son99,Dolgov82,Dolgov08,Combescot97} 
dealing with this subject that evolved since then departs from an assumption dating back to the seminal work of 
Herz in the 1970's\cite{Hertz76}. This involves the nature of the ultraviolet: at some relatively short time scale where the electron system has closely
approached a Fermi-liquid the influence of the critical order parameter fluctuations become noticable. The Fermi surface and Fermi energy 
of this quasiparticle system can then be used as building blocks together with the bosonic field theory describing the critical order parameter
fluctuations to construct a perturbative framework dealing with the coupling between these fermionic- and bosonic sectors. The lowest order
effect of this coupling is that the fermi gas of quasiparticles acts as a heat bath damping the bosonic order parameter fluctuations, with the
effect that the effective space-time dimensionality of the bosonic field theory exceeds the upper  critical dimension. These dressed order
parameter fluctuations than 'back react' on the quasiparticle system  causing  'singular' interactions in the Cooper channel, yielding in turn a
rational for a generic  'high Tc' superconductivity at QCP's.   

The crucial assumption in this 'Herz philosophy'  is that the fermion physics is eventually controlled by the Fermi gas. In the cases of 
empirical interest it is generally agreed that in the UV the interaction energies are much larger than the bare kinetic energies, while 
there is no obvious signature in the experiments for a renormalization flow that brings the system close to a weakly interacting fermion
gas before entering the singular 'Herz' critical regime. From the theoretical side, the introduction of this UV Fermi gas can be viewed as
an intuitive leap. The only truly fermionic state of matter  that is understood mathematically is the Fermi gas and its perturbative 'derivative'  
(the Fermi liquid): the fermion sign problem makes it impossible to address fermionic matter in general mathematical terms\cite{Troyer05}. However,
very recently the 'grib of the Fermi-gas' has started to loosen specifically in the context of fermionic critical matter. A first step in this direction
is the demonstration of proof of principle that truly critical fermionic states of matter can exist that have no knowledge whatever of the 
statistical Fermi energy scale: the fermionic Feynman backflow wavefunction Ansatz\cite{Kruger08}.  The substantive development is the recent work
addressing fermion physics using the string theoretical AdS/CFT correspondence. It appears that this duality between quantum field 
theory and gravitational physics is capable of describing Fermi-liquids that emerge from a manifestly strongly interacting, critical ultraviolet\cite{Schalm09}.
In another implementation, one finds an IR physics describing 'near' Fermi-liquids characterized by 'critical' Fermi surfaces\cite{Senthil08} controlled by
an emergent conformal symmetry implying the absence of energy scales like the  Fermi-energy\cite{Liu0901,Liu0902}.

This lengthy consideration is required to motivate the subject of this paper: a phenomenological scaling theory for a Bardeen-Cooper-Schrieffer
 (BCS) type  superconductivity starting from the postulate that the normal state is not a Fermi-liquid, but instead  a truly conformal 
fermionic state of matter. With 'BCS type' we mean the following: we assume as in BCS that besides the electron system a bosonic modes 
are present that cause attractive electron-electron interactions. This 'glue' is retarded in the sense that the characteristic energy scale of 
this external bosonic system $\omega_B$ is small as compared to the ultraviolet cut-off  scale of the quantum critical fermion system $\omega_c$ . 
Having a small Migdal parameter, the glue-electron vertex corrections can then be ignored 
and the  the effects of the glue are described in terms of the Migdal-Eliashberg  time dependent mean field theory, reducing to the static BCS mean 
field theory in the weak coupling limit\cite{Schrieffer71}.  All information coming from the electron system that is required for the pairing instability 
is encapsulated in the electronic pair susceptibility. Instead of using the Fermi gas pair susceptibility (as in conventional BCS),  we rely on the fact
that conformal invariance fixes the analytical form of this response function in terms of two free parameters: an overall UV cut-off scale ($T_0$) and
the  anomalous scaling dimension of the pair susceptibility, expressed in a dynamical critical exponent $z$ and correlation function exponent 
$\eta_p$.  The outcome is a scaling theory for superconductivity that is in essence very simple; much of the technical considerations that follow 
are dealing with details associated with modeling accurately the effects of the  breaking of conformal invariance by  temperature and the 
superconducting instability.  This theory is however surprisingly economical in yielding phenomenological insights. Conventional BCS appears
as a special 'marginal' case, and our main result is the generalized gap equation, Eq. (\ref{qcgap}). The surprise it reveals is the role of retardation:
when the Migdal parameter $\omega_B/\omega_c$ is small (where the mathematical control is best) we find at small coupling constants 
$\tilde{\lambda}$ a completely different behavior compared to conventional BCS: the gap magnitude $\Delta$ becomes similar to the glue
energy $\omega_B$. To illustrate the case with numbers, a moderate coupling to phonons like $\tilde{\lambda} = 0.3$ with a  frequency
$\omega_B = 50$ meV will yield rather independently of scaling dimensions a gap of 40 $meV$ and  a $T_c$ of 100 Kelvin or so: these
are numbers of relevance to cuprate superconductors! 

The theory has more in store. Incorporating the motive that on both sides of the quantum critical point heavy Fermi liquids emerge from
the quantum critical metal as in the heavy fermion systems, we show that the superconducting 'dome' surrounding the quantum critical 
point emerges naturally without changing the coupling to the bosonic glue. The form of this dome is governed by the correlation length,
but we find via the pair susceptibility a direct relation with the effective mass of the quasiparticles of the 
Fermi-liquids. Last but not least, we analyze the orbital limiting upper critical magnetic field, finding out that pending the value of the 
dynamical critical exponent it can diverge very rapidly upon approaching the QCP, offering an explanation for the observations in the
ferromagnetic URhGe heavy fermion superconductor\cite{Levy07}.

The scaling phenomenology we present here is simple and obvious, but it appears to be overlooked so far.  Earlier work by Balatsky\cite{Balatsky93}, Sudbo\cite{Sudbo95}  
and Yin and Chakravarty\cite{Chakravarty96}
is similar in spirit but yet quite different. These authors 
depart from a Luttinger liquid type single particle propagators to compute the pair susceptibility from the bare fermion particle-particle loop.
Although this leads to pair susceptibility similar (although not identical) to ours, it is conceptually misleading since in any non Fermi-liquid, there is no such simple 
relation between two-point and four-point correlators. This is particularly well understood for conformal field theories: for the higher
dimensional cases the AdS/CFT correspondence demonstrates that two point CFT correlators are determined by kinematics in AdS while
the four- and higher point correlators require a tree level computation\cite{Muck98,Freedman99,Hoker98,Liu98,Hoker9802,Hoker99,Schalm98,Schalm99}. 
More serious for the phenomenology, this older
work ignores the role played by retardation; it is a-priori unclear whether one can construct a mathematically controlled
 scaling theory for BCS  without the help of a small Migdal parameter.
 
 The remainder of this paper is organized as follows. In section II we review a somewhat unfamiliar formulation of the classic BCS  
 theory that makes very explicit the role of the pair susceptibility. We then introduce the scaling forms for the pair susceptibilities    
 as follow from conformal invariance. By crudely treating the modifications in the pair susceptibility at low energies associated 
 with  the presence of the pair condensate we obtain the new gap equation Eq. (\ref{qcgap}). This catches already the essence of
 the BCS superconductivity of quantum critical metals and we discuss its implications in detail. In section III we focus in on intricacies
 associated with determining the transition temperature. Conformal invariance is now broken and one needs to know the scaling
 functions  in some detail. We use the exact results of 1+1 dimensional conformal field theory as a model to address these matters.
 In section IV we turn to the harder problem of modeling the crossover from the large energy critical pair susceptibility
 to the low energy, zero temperature infrared that is governed by conventional  Bogoliubov fermions, as needed to devise a more
 accurate zero temperature gap equation. The casual reader might
 want to skip both sections. The moral is that information on the cross-over behavior of  the pair susceptibility is required that is 
beyond simple scaling considerations to address what happens when  the conformal invariance is broken either by temperature
 (as of relevance to the value of $T_c$) or by the presence of the BCS condensate (of relevance for the zero temperature gap). 
 The conclusion will be that although the gross behaviors are not affected, it appears to be impossible to compute numbers like
 the gap to $T_c$ ratio accurately since these are sensitive to the details of the cross-over behaviors. In section V we explore
 the theory away from the critical point, assuming that cross-overs follow to heavy Fermi-liquids, where we address the 
 origin of the superconducting dome.  Finally, in section VI we address  the scaling behavior of the orbital limited upper critical field.

\section{BCS theory and the scaling of the pair susceptibility.}     
      
Let us first revisit the backbone of Migdal-Eliashberg theory.  We need a formulation that is avoiding the explicit references to
the Fermi gas of the text book formulation, but it is of course well known how to accomplish this. 
Under the condition of strong retardation and small couplings, the effects of the glue are completely enumerated by the gap equation\cite{Allen80}
ignoring angular momentum channels ($s$,$d$ waves, etcetera) for the time being,
\begin{equation}
 1-g\chi'_{\rm ret}({\vec q}=0,\omega=0,\Delta,T)=0,
 \label{gapeq}
\end{equation}
where $g$ is the effective coupling strength of the glue, while $\chi'_{\rm ret}$ is the zero frequency value of the real part of the 
retarded pair susceptibility at a temperature $T$ in the presence of the gap $\Delta$. This effective $\chi'_{\rm ret}$ also incorporates 
the effects of retardation. The textbooks with their focus on non-interacting electrons accomplish this in a rather indirect way, by 
putting constraints on momentum integrations. Retardation is however about time scales and the general way to incorporate retardation
is by computing $\chi'_{\rm ret}$ by employing the Kramers-Kronig relation starting from the imaginary part of the full electronic pair susceptibility $\chi_p''$.
For a glue characterized by a single frequency $\omega_B$,
\begin{equation}
\chi'_{\rm ret}(\omega=0)=2 {\cal P} \int_0^{2\omega_B}d\omega' \frac{\chi_p''(\omega')}{\omega'}.
\label{retard}
\end{equation} 
with the full pair susceptibility given by the Kubo formula, 
\begin{equation}
\chi_p({\vec q},\omega)=-i\int_0^{\infty}dte^{i(\omega+i\eta)t}\left\langle [b^{\dagger}({\vec q},0),b({\vec q},t)] \right\rangle,
 \end{equation} associated with the pair operator $b^{\dagger}({\vec q},t)={\sum_{\vec k}}
c^{\dagger}_{{\vec k}+\frac{\vec q}{2},\uparrow}(t)c^{\dagger}_{-{\vec k}+\frac{\vec q}{2},\downarrow}(t)$. 

In the case of conventional superconductors the normal state is a Fermi-liquid, formed from (nearly) non-interacting quasiparticles. One can
get away with a 'bare fermion loop' pair susceptibility. The specialty of this pair susceptibility is that its imaginary part is frequency independent 
at zero temperature. It extends up to the Fermi energy of the Fermi-liquid and from the unitary condition,
\begin{equation}
\int_0^{\infty}\chi_p''(\omega)d\omega=1 
\label{norm}
\end{equation}
it follows that   at zero temperature $\chi''(\omega) = N_0 = 1/(2E_F)$. In logarithmic accuracy the 
gap enters as the low frequency cut-off in Eq. (\ref{retard}) such that, 
\begin{equation}
\chi'_{\rm ret} (\omega=0,\Delta,T=0) = \int_{\Delta}^{2\omega_B}\frac{d\omega'}   { E_F \omega'}
= \frac{1}{E_F} \log \frac{2 \omega_B}{\Delta},
\end{equation}
and from Eq. (\ref{gapeq}) the famous BCS gap equation follows: $\Delta = 2\omega_B e^{-1/\lambda}$, where $\lambda = g/E_F$.
 
This formulation of BCS has the benefit that it makes very explicit that all the information on the electron system 
required for the understanding of the pairing instability is encoded in the pair susceptibility. This is in turn a
bosonic response function of the electron system since it involves the response of two fermions, much like 
the dynamical susceptibilities associated with charge- or spin densities. In addition one needs the fact that the pair density
is a non-conserved quantity, in the same sense as a staggered magnetization.  When the quantum system is conformal 
(i.e. the zero temperature quantum critical metal) the analytical form of the  dynamical pair susceptibility is fixed at zero
temperature by the requirement  of invariance under scale transformations\cite{Sachdev99},
\begin{equation}
 \chi(\omega)=  \lim_{\delta \rightarrow 0} Z''(-(\omega+i\delta)^2)^{-\frac{2-\eta_p}{2z}},
 \label{zeroTconf}
\end{equation}
as determined by the a-priori unknown unknown exponents $\eta_p$ and $z$, the anomalous scaling dimension of the pair operator
and the dynamical critical exponent, respectively. The normalization constant $Z''$ is via the unitarity condition Eq.(\ref{norm}) 
determined by the UV cut-off scale $\omega_c$. Because we invoke a small Migdal parameter we are interested in the 'deep
infrared' of the theory that is not very sensitive to the precise choice of this UV energy scale. A reasonable choice is the energy 
where the thermal de Broglie wavelength becomes of order of the electron separation, i.e. the Fermi energy of an equivalent system
of non-interacting electrons. Defining $\alpha_p = \frac{2-\eta_p}{z}$ and using Eq. (\ref{norm}) with the cut-off scale $\omega_c$, we find, \begin{equation}
Z'' = \frac{1-\alpha_p}{\sin(\frac{\pi}{2}\alpha_p)}\frac{1}{\omega_c^{1-\alpha_p}},
\label{normZdouble}
\end{equation}
observing that $\alpha_p < 1$ in order for this function to be normalizable: this is the well known unitary bound on the operator dimensions. The real and imaginary parts of the zero temperature critical pair susceptibility are related by a phase angle $\frac{\pi}{2}\alpha_p$,
\begin{equation}
 \chi(\omega)= \frac{Z''}{\omega^{\alpha_p}}\left( \cos(\frac{\pi}{2}\alpha_p)+i\sin(\frac{\pi}{2}\alpha_p)\right).
 \end{equation}
 
 \begin{figure}
\begin{center}
\includegraphics[width=0.43\linewidth]{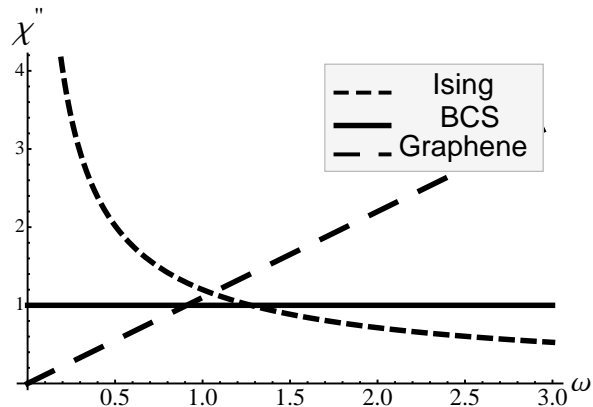}
\end{center}
\caption{Illustration of the imaginary part of the pair susceptibility, comparing the relevant (Ising class),  marginal (BCS case) and irrelevant (graphene class). The scaling exponent $\alpha_p = \frac{2-\eta_p}{z}$ is respectively $0<\alpha_p<1, \alpha_p=0, \alpha_p<0$. For the Ising class, the magnitude of the imaginary part of the pair susceptibility becomes larger and larger as one lowers the frequency. For the BCS case, the magnitude stays constant as the frequency is changed. For the graphene class, the magnitude decreases to zero in the low frequency infrared region. }
\end{figure}

According to general conformal wisdoms, the pair operator is called irrelevant when $\alpha_p < 0$ such that $\chi"$ increases with frequency,
relevant when $0 < \alpha_p < 1$ when $\chi''$ decreases with frequency and marginal when $\alpha_p =0$, such that $\chi''_p$ is frequency
independent, see Fig 1. From this scaling perspective, the Fermi liquid  pair operator is
just the special marginal case,  and the BCS superconductor with its logarithmically running coupling constant falls quite literally in the same
category as the asymptotically free quantum chromo dynamics in 3+1D and the Kondo effect. Another familiar case is the pair susceptibility 
derived from the 'Dirac fermions' of graphene\cite{Uchoa07,Kopnin08} and transition metal dichalcogenides\cite{Neto01,Uchoa05} characterized by 
$\alpha_p = -1$: in this 'irrelevant case' one needs a finite glue interaction to satisfy the instability criterium.  

The scaling behavior of the free fermion case is special and the pair operator in a general conformal fermionic state can be characterized by a 
scaling dimension that is any real number smaller than one.  
 Obviously, the interesting case is  the relevant one where $\alpha_p > 0$  (Fig.1).  Let us here consider the zero temperature gap equation. 
In Eq. (\ref{zeroTconf}) we have already fully specified $\chi''_p$ in the critical state. However, due to
the zero temperature condensate the scale invariance is broken and the low frequency part of $\chi''_p$ will now be dominated by an emergent 
BCS spectrum including a $s-$ or $d-$wave gap, Bogoliubov fermions and so forth. 
This will be discussed in detail in  section V. Let us here introduce the gap in the BCS style by just assuming that the imaginary part of the pair susceptibility
vanishes at energies less than $\Delta$.  Under this assumption the gap equation becomes,
\begin{equation}
\label{qcgap00}
1-2g\int_{\Delta}^{2\omega_B} \frac{d \omega}{\omega} \frac{Z''\sin((\pi/2)\alpha_p)}{\omega^{(2-\eta_p)/z}}=0,
\end{equation}
evaluating the integral this becomes our 'quantum critical gap equation' ,
\begin{equation}
\label{qcgap}
\Delta = 2\omega_B  \left({1+\frac{1}{\tilde{\lambda}}\left(\frac{2\omega_B}{\omega_c} \right)^{\alpha_p} } \right) ^{-\frac{1}{\alpha_p}},
\end{equation}
with 
\begin{equation}
\label{lambdatilde}
\tilde{\lambda}= 2\lambda\frac{1-\alpha_p}{\alpha_p} ,
\end{equation}
and $\lambda\equiv g/\omega_c$. The numerator $(1-\alpha_p)$ in $\tilde{\lambda}$ comes from the normalization constant $Z''$, while the denominator $\alpha_p$ from integrating over $\omega$. Notice that $\lambda$ has the same meaning of a
conventional, say, dimensionless electron-phonon coupling constant. The dimensionful coupling constant $g$ parametrizes the interaction
strength between microscopic electrons and -lattice vibrations, and $\omega_c$ has the same status as the Fermi-energy in a conventional
metal as the energy scale  that is required  to balance $g$. We argued earlier that $\omega_c$ is of order of the bare Fermi energy and 
therefore it make sense to use here values for e.g.  the electron-phonon coupling constant as quoted in the LDA literature. Notice however that for a given 
$\lambda$ the effective coupling constant $\tilde{\lambda}$ that appears in Eq. (\ref{qcgap}) is decreasing when $\alpha_p$ is becoming more relevant, i.e. when
$\alpha_p \rightarrow 1$.
From the frequency integral $\int d\omega \omega^{-(1+\alpha_p)}$, one would anticipate that the gap would increase for a more relevant pair susceptibility. The unitary condition imposes however an extra condition on the pair susceptibility. These two compensating effects lead to the important result that the gap is rather sensitive to the relevancy of the pair susceptibility.  All what really matters is whether the pair susceptibility is relevant rather than marginal or irrelevant, and the degree of the relevancy is remarkably unimportant.

\begin{figure}
\begin{centering}
\includegraphics[width=1.0\linewidth]{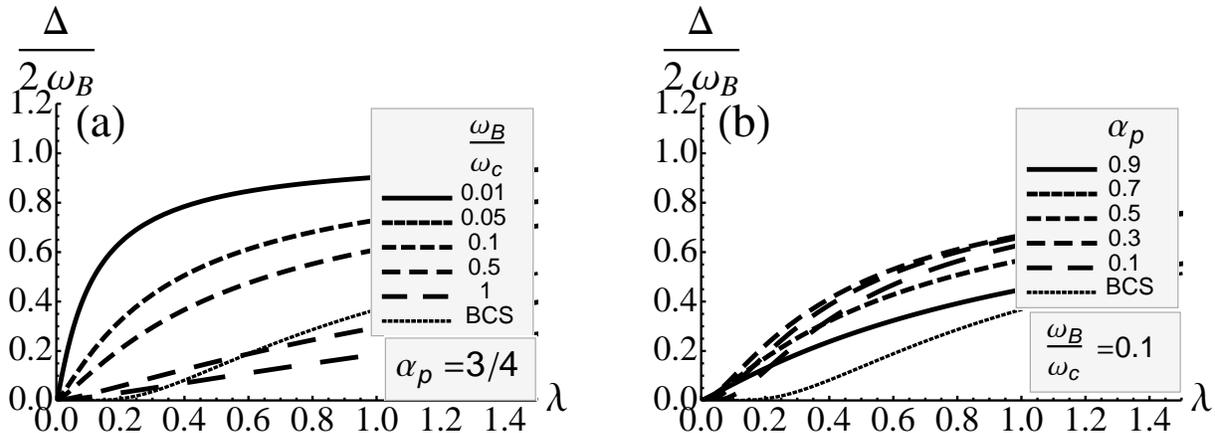}
\end{centering}
\caption{(a)The ratio of gap to retardation frequency $\Delta/(2\omega_B)$ as a function of glue strength $\lambda$ for various retardation ranges $\omega_B/\omega_c$ with fixed scaling dimension $\alpha_p=3/4$. (b)The same plot, but with fixed retardation $\omega_B/\omega_c=0.1$ and various scaling dimensions $\alpha_p$. The dotted lines are the standard BCS result.}
\end{figure}

Eq.(\ref{qcgap}) is a quite different gap equation than the BCS one with its exponential dependence on the coupling $\lambda$.  The multiplicative structure associated with the Fermi-liquid is scaling wise quite
special, while Eq. (\ref{qcgap}) reflects directly  the algebraic structure rooted in scale invariance. The 
surprise is that retardation acts quite differently when power laws are ruling. In Fig. (1) we show the dependence of the $\Delta/\omega_B$ ratio on the coupling constant $\lambda$, both for different
Migdal parameters and fixed $\alpha_p$, as well as for various scaling dimensions and the Migdal parameter fixed. The comparison with the BCS result shows that drastic changes happen already for small scaling dimensions $\alpha_p$ especially in the  small $\lambda$ regime. Our equation actually predicts  that the gap to glue frequency ratio becomes of order
one alrady for couplings that are as small as $\lambda =0.1$  when the Migdal parameter is small. To place this in the context of high Tc superconductivity, let us assume that the pairing glue in the cuprates
is entirely rooted in the 'glue peak' at $\omega_B \sim 50$meV that is consistently detected photoemission, tunneling spectroscopy  and optical spectroscopy\cite{Davis06,Shen03,Heumen09} . 
The electronic cut-off in the
cuprates is likely of order $\omega_c = 0.5$ eV such that the Migdal parameter $\omega_B/\omega_c \simeq 0.1$. A typical gap value is $40$ meV and we read off Fig. 1 that we need $\lambda = 0.45$
or $0.43$ for $\alpha_p = 3/4, 1/4$  while using the BCS equation $\lambda = 1.1$! Taking this serious implies  that in principle one needs no more than a standard electron-phonon coupling to explain superconductivity
at a high temperature in cuprate superconductors. Of course this does not solve the problem: although one gets a high Tc for free it still remains in the dark how to form a fermionic quantum critical state 
with a high cut-off energy, characterized by a relevant pair susceptibility. 

Eq.(\ref{qcgap}) is also very different from the gap equations obtained in the previous attempts to apply scaling theory to superconducting transition  by Balatsky\cite{Balatsky93}, Sudbo\cite{Sudbo95}  and Yin and Chakravarty\cite{Chakravarty96}. A crucial property of their results is that even in the relevant case one needs to exceed a critical value for $\lambda$ to find a superconducting instability. The present scaling theory is in this regard a more natural generalization of BCS theory,  where the standard BCS is just the 'marginal end' of the relevant regime where the Cooper instability cannot be avoided for attractive interactions.
The previous approaches \cite{Balatsky93,Sudbo95,Chakravarty96} start by considering the single particle spectral function, generalizing its analytic structure from simple poles to branch cuts. This way of thinking stems from the Fermi-liquid type assumption that the single particle Green's function is the only primary operator of the system, and all the higher point functions are secondary operators, to be determined by the single particle Green's function. But for critical systems, such assumptions are generally not to satisfied. It is well known for example from the AdS/CFT correspondence, that the four-point functions of strongly interacting conformal fields are much more complex than the combination of  two-point functions\cite{Muck98,Freedman99,Hoker98,Liu98,Hoker9802,Hoker99,Schalm98,Schalm99}. Our basic assumption is that the pair susceptibility is by itself a primary operator  subjected to conformal invariance which
is the most divergent operator at the critical point.

\section{Determining the transition temperature.}

Let us now turn to finite temperatures. A complicating fact is that temperature breaks conformal
invariance, since in the euclidean formulation of the field theory its effect is that the periodic imaginary
time acquires a finite compactification radius $R_{\tau} = \hbar / k_B T$. The pair susceptibility therefore
acquires the finite size scaling form\cite{Sachdev99}
 \begin{equation}
\chi(\omega)\equiv\chi({\vec q}=0,\omega)=ZT^{-(2-\eta_p)/z}\Phi\left( \frac{\omega}{T}\right),
\end{equation}
where $\Phi$ is a universal scaling function and $Z$ is a UV renormalization constant, while $\eta_p$ and $z$ are the anomalous scaling dimension of the pair operator and the dynamical critical exponent, 
respectively. At zero temperature this turns into the banch cut as shown in Eq.(\ref{zeroTconf}),  while in the opposite high temperature or hydrodynamical regime ($\hbar \omega << k_B T$) it takes the form\cite{Sachdev99}
\begin{equation}
\chi(\omega)=Z'T^{-(2-\eta_p)/z}\frac{1}{1-i\omega\tau_{rel}},
\label{highTconf}
\end{equation}
 where $\tau_{rel}\approx\hbar/k_BT$. 
The crossover from the hydrodynamical- (Eq. \ref{zeroTconf}) to  the high frequency coherent regime(Eq. \ref{highTconf}) occurs at an energy $\sim k_BT$.  The superconducting transition
 temperature is now determined by the gap equation through $1 - g \chi'_{\rm ret} (k_B T_c)= 0$. The problem is that $\chi'_{ret}$ is via the Kramers-Kronig
 transformation largely set by the cross-over regime in $\chi''_p$.  One needs the full solutions of the CFT's to determine the detailed form of  $\Phi$ in this crossover regime and these are not
 available in higher dimensions.
 
 In 1+1D these are however completely determined by conformal invariance, and for our present purposes these results might well represent 
 a reasonable model since the gap equation is only sensitive to rather generic features of the cross-over behavior. Given the exponents $\eta_p$
 and $z$, the exact result for the finite temperature $\chi''$ in 1+1D is well known and can be easily derived from the universal two-point correlator at an imaginary time $\tau$\cite{Sachdev99}
 \begin{equation}
 C(x,\tau)={\tilde Z}\frac{T^{2s}}{\left(\sin[\pi T (\tau-ix)]\sin[\pi T (\tau+ix)]\right)^s}, 
 \end{equation}
 with $1-2s=\frac{2-\eta_p}{2z}$. The  analytical continuation to real time $\tau\to it$ yields the real time two-point correlation function
  \begin{equation}
 C(x,t)={\tilde Z}\frac{T^{2s}}{\left(i\sinh[\pi T (t-x)]i\sinh[\pi T (t+x)]\right)^s},
 \end{equation}
 with a Fourier transform corresponding to the dynamic structure factor
 \begin{equation}
 S(k,\omega)=\int_{-\infty}^{\infty}dx\int_{-\infty}^{\infty}dt C(x,t)e^{-i(kx-\omega t)}.
 \end{equation}
A convenient way to perform  the Fourier transform is by factorizing $C(x,t)$ into left-moving and right-moving modes, $C(x,t)=C_{-}(t-x)C_{+}(t+x)$, to subsequently integrate over $t\pm x$. The result is
  \begin{equation}
S(k,\omega)=Ze^{\frac{\omega}{2T}}\frac{1}{T^{2(1-2s)}}B(s+i \frac{\omega+k}{4 \pi T},s-i \frac{\omega+k}{4 \pi T})B(s+i \frac{\omega-k}{4 \pi T},s-i \frac{\omega-k}{4 \pi T}),
\end{equation}
where $B$ is the beta function, and the overall numerical coefficient $Z=2^{4s-3}\pi^{2(s-1)}{\tilde Z}$. The fluctuation-dissipation theorem 
\begin{equation}
S(k,\omega)=\frac{2}{1-e^{-\omega/T}}\chi''(k,w)
\end{equation}
 then yields the imaginary part of the pair susceptibility,
 \begin{equation}
 \label{imchifiniteT}
\chi''(k,\omega)=Z\frac{\sinh(\frac{\omega}{2T})}{T^{2(1-2s)}}B(s+i \frac{\omega+k}{4 \pi T},s-i \frac{\omega+k}{4 \pi T})B(s+i \frac{\omega-k}{4 \pi T},s-i \frac{\omega-k}{4 \pi T}),
\end{equation}
The temperature and frequency dependencies of this function for $k=0$ are illustrated in Fig.(3). Indeed $\chi''(\omega)\to 0$ in a linear fashion with $\omega$ with a slope set by $1/T$, while for $\omega >>T$ the 
temperature dependence drops out, recovering the power law. The crossover occurs at  $\omega\simeq 2k_B T/\hbar$ where $\chi''(\omega)$ has a maximum.
In the absence of retardation, the real part can be computed from the Kramers-Kronig transform, 
\begin{eqnarray}
\chi'(k,\omega)=\frac{Z'}{T^{2(1-2s)}}&\left( \frac{-i\pi}{s-i \frac{\omega+k}{4 \pi T}}  \frac{\sin(2s\pi-\frac{ik}{2T})}{\sinh(\frac{k}{2T})}\frac{\Gamma(2s)\Gamma(2s-\frac{ik}{2\pi T})}{\Gamma(1-\frac{ik}{2\pi T})} \:
{}_3F_2 (2s,s-i \frac{\omega+k}{4 \pi T},2s-\frac{ik}{2\pi T};1+s-i \frac{\omega+k}{4 \pi T},1-\frac{ik}{2\pi T};1)\right.&\nonumber\\  &+\left.(k\rightarrow -k)  \right)&,
\end{eqnarray} 
where $F$ is the generalized hypergeometric function. We did not manage to obtain an analytic form for the real part when retardation is included, and we use numerical results instead. 
\begin{figure}
\begin{centering}
\includegraphics[width=1.0\linewidth]{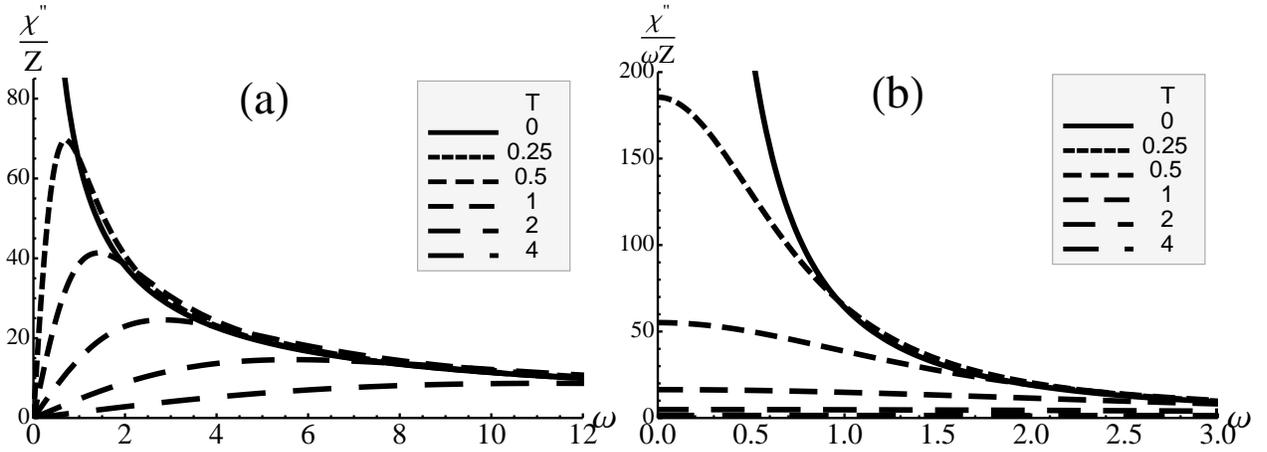}
\end{centering}
\caption{(a)Illustration of the imaginary part of the pair susceptibility $\chi''(k=0,\omega,T)$ divided by the overall numerical factor $Z$, as a function of frequency $\omega$ for various temperatures. Here we've chosen $\alpha_p=3/4$, so $s=5/16$. (b) The same plot, but $\chi''$ is further divided by $\omega$. At zero temperature one has the power law scaling form. At finite temperature $\chi''(\omega)$ goes to zero, as $\omega$ goes to zero ($\chi''(\omega)/\omega\to {\rm constant}$, as $\omega\to 0$), and approaches the same power law behavior at large frequency. As one increases temperature, the maximum of $\chi''(\omega)$ goes down, and the corresponding $\omega_{max}$ shifts to larger frequency.}
\end{figure}

When temperature goes to zero the limiting form of the  beta function becomes,
\begin{equation}
\lim_{u\to\infty}B(s+iu,s-iu)=\frac{2\pi}{\Gamma(2s)}e^{-\pi u}u^{2s-1},
\end{equation}
and the imaginary part of the pair susceptibility Eq. (\ref{imchifiniteT}) acquires the power law form
\begin{equation}
\chi''(\omega)=\frac{2\pi^2(4\pi)^{\alpha_p}}{[\Gamma(2s)]^2}Z\frac{1}{\omega^{\alpha_p}}.
\end{equation}
Comparing this with Eq.(\ref{normZdouble}) yields the normalization factor in terms of  the cut-off scale
\begin{equation}
\label{ZfiniteT}
Z=\frac{[\Gamma(2s)]^2(1-\alpha_p)}{2\pi^2(4\pi)^{\alpha_p}\omega_c^{1-\alpha_p}}.
\end{equation}

Combining Eq.'s (\ref{gapeq}),(\ref{retard}),(\ref{imchifiniteT}),(\ref{ZfiniteT}), we obtain the equation determining the critical temperature,
 \begin{equation}
\label{tc1}
1-{\cal C}' \lambda\left( \frac{2\omega_B}{\omega_c}\right)^{-\alpha_p}  \left( \frac{T_c}{2\omega_B}\right)^{-\alpha_p}{\cal F}\left( \frac{2\omega_B}{T_c}\right) =0,
\end{equation}
where
\begin{equation}
{\cal F}(y)=\int_0^{y}\frac{dx}{x}\sinh(\frac{x}{2})\left( {\rm B}(s+i \frac{x}{4\pi},s-i \frac{x}{4\pi})\right) ^2,
\end{equation}
and $x=\omega/T$. The overall coefficient is
 \begin{equation}
 {\cal C}'=\frac{[\Gamma(2s)]^2(1-\alpha_p)}{\pi^2(4\pi)^{\alpha_p}}.
 \end{equation}

 \begin{figure}
\begin{centering}
\includegraphics[width=1.0\linewidth]{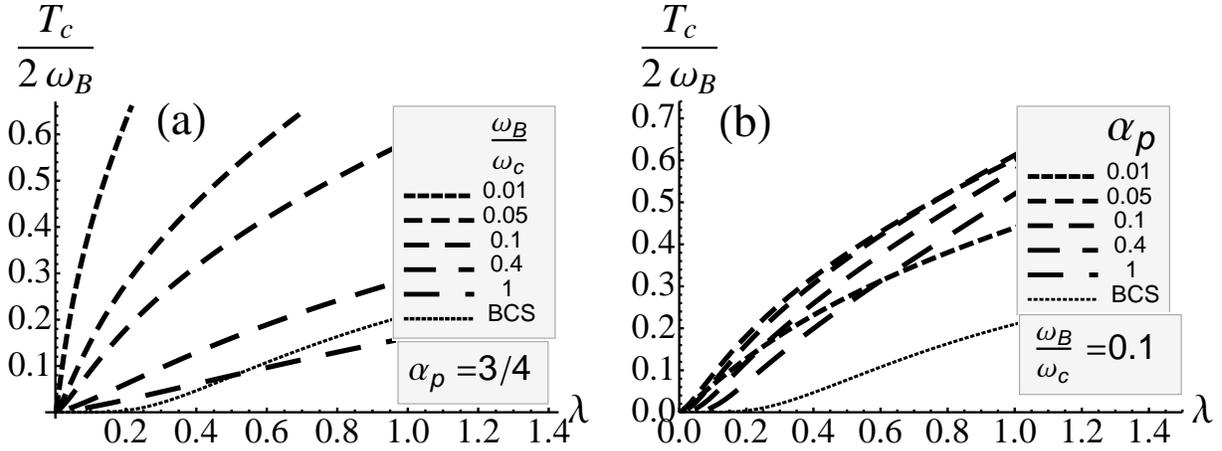}
\end{centering}
\caption{(a)The ratio of transition temperature to retardation frequency $T_c/(2\omega_B)$ as a function of glue strength $\lambda$ for various retardation ranges $\omega_B/\omega_c$, with  scaling dimension fixed $\alpha_p=3/4$. (b)The same plot, but fix the retardation $\omega_B/\omega_c=0.1$ while varying the scaling dimensions $\alpha_p$. The dotted lines are the standard BCS result. The magnitude and dependence on glue strength and retardation are all similar to those of the gap.}
\end{figure}

We plot in Fig.(4) the ratio of $T_c$ to retardation frequency as function of glue strength, retardation and the scaling dimensions. One infers that the behavior of $T_c$ is very similar to that of the zero temperature gap, plotted in Fig. (2). We observe that they are of the same order of magnitude $T_c\sim \Delta$, and this can be understood from the behavior of $\chi''/\omega$ plotted in Fig.(3b). Since the large frequency behavior of $\chi''(\omega)/\omega$'s are the same for different temperatures, all what matters is the low frequency part. The gap imposes a cut-off for the zero temperature $\chi''(\omega)/\omega$, and its value is determined such that the area under this curve including the low frequency cut-off, is the same as the area under the curve corresponding to $T_c$ without a cut-off: by inspecting Fig.(3b) one infers directly that the gap and $T_c$ will be of the same order. The same logic is actually at work in the standard BCS case. The finite temperature Fermi gas susceptibility is $\chi''(\omega)=\frac{1}{2E_F}\tanh(\frac{1}{4}\beta\omega)$ \cite{Allen80}, and the familiar $T_c$ equation follows,
\begin{equation}
1-\lambda\int_0^{2\omega_B}\frac{d\omega}{\omega}\tanh(\frac{1}{4}\beta\omega)=0,
\end{equation}
such that
$T_c\simeq1.14\omega_B e^{-1/\lambda}$,  of the same order as the BCS gap $\Delta = 2\omega_B e^{-1/\lambda}$. Now the effect of temperature is encoded in the $\tanh$ function. Although the Fermi-gas is not truly
conformal, It is easy to check that this 'fermionic' $\tanh$ factor adds a temperature dependence to the $\chi''$ that is nearly indistinguishable from what one obtains from the truly conformal marginal case that one obtains
by setting $s=1/2$ in Eq. (\ref{imchifiniteT}).  

We notice that conformal invariance imposes severe constraints on the finite temperature behavior of the pair susceptibility, thereby simplifying  the calculation of  $T_c$. In the  $1+1$-dimensional 'model' 
nearly everything is fixed by conformal invariance.  The only free parameters that enter the calculation are the scaling dimension $\alpha_p$,  the cut-off scale $\omega_c$ and the glue quantities.  As we will
now argue the situation is actually much less straightforward for the zero temperature gap because this involves a detailed  knowledge of the crossover to the physics of the superconductor ruling the low energy
realms.

\section{The gap equation: gluing a quantum critical metal to a BCS superconductor.}

It is part of our postulate that when superconductivity sets in BCS 'normalcy' returns at low energies in the form 
of the sharp Bogoliubov fermions and so forth. Regardless the critical nature of the normal state, the scale invariance
gets broken by the instability where the charge 2e Cooper pairs form, and this stable fixed point also dictates the nature
of the low lying excitations. However, we are dealing with the same basic problem as in the previous section: in the absence
of a solution to the full, unknown theory it is impossible to address the precise nature of the cross-over regime between the
BCS scaling limit and the critical state at high energy. This information is however required to further improve the gap equation
Eq. (\ref{qcgap}) of section II that was derived by crudely modeling $\chi''$ in the presence of the superconducting condensate. 

\begin{figure}
\begin{centering}
\includegraphics[width=1.0\linewidth]{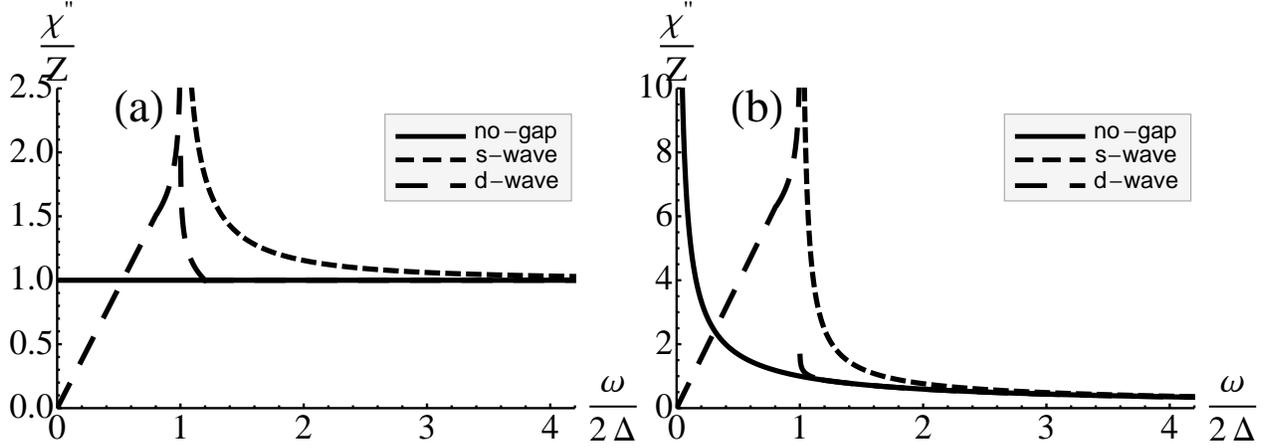}
\end{centering}
\caption{Illustration of the imaginary part of the pair susceptibility without a gap and in the presence of an $s$- and $d$- wave gap, for (a) the BCS case, and (b) the quantum critical case (here we've plotted using the parameter $\alpha_p=3/4$). In the absence of gap, $\chi''$ is a constant (for BCS) or has a simple power law behavior (for critical fermions). In the presence of a s-wave gap, the states below the gap are gapped out and there is a power law singularity right above the gap. When there is a d-wave gap, the low frequency part (way below the gap) is governed by a Dirac cone structure, thus a linear susceptibility,  while near the gap a van Hove singularity is at work, leading to logarithmic divergences on both sides. The high frequency region for both s- and d-wave gap goes over to the case without a gap.}
\end{figure}

So much is clear that the crossover scale itself is set by the gap magnitude $\Delta$. However, assuming that this affair has
dealings with e.g. optimally doped cuprate superconductors, we can rest on experimental information: in optimally doped
cuprates at low temperatures the coherent Bogoliubov fermions persist as bound states all the way to the gap maximum.
Up to these energies it is therefore reasonable to assume that $\chi''_p$ is determined by the bare fermion loops, and this
regime has to be smoothly connected to the branch cut form of the $\chi''_p$ at higher energies. This implies that 
the standard BCS gap singularities have to be incorporated in our zero temperature pair susceptibility. As a final requirement,
the pair susceptibility has to stay normalized according to Eq. (\ref{norm}), which significantly limits the modelling freedom.    

\begin{figure}
\begin{centering}
\includegraphics[width=1.0\linewidth]{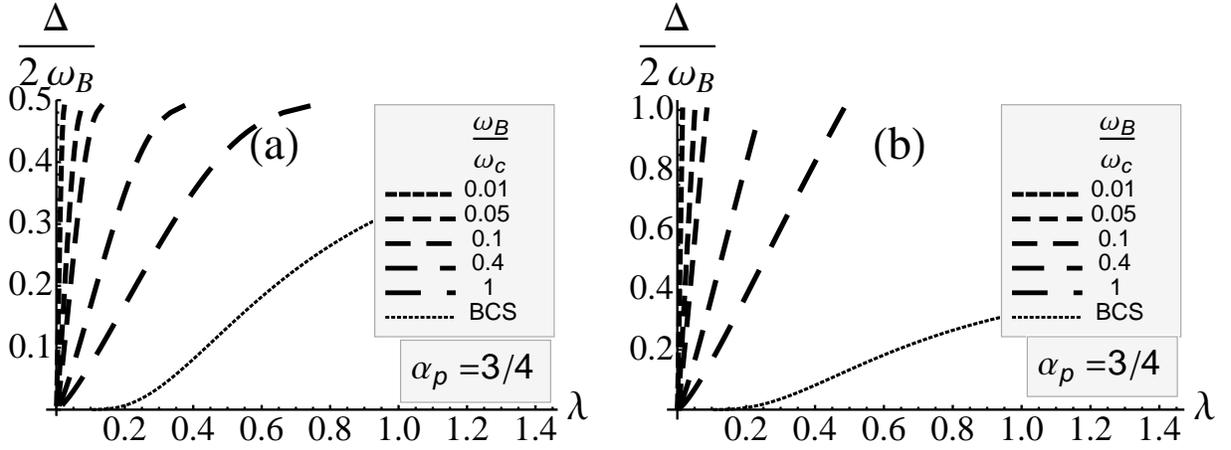}
\end{centering}
\caption{The ratio of the gap to retardation $\Delta/(2\omega_B)$ as a function of the glue strength $\lambda$, for various retardation ranges, with (a) a s-wave gap and (b) a d-wave gap. Here we've chosen $\alpha_p=3/4$. The dotted lines are the standard BCS result. The dependence on glue strength and and retardation is similar but the magnitude of the gap is much enhanced compared to the previous treatment of gap as a simple IR cutoff. The d-wave case is enhanced even more. }
\end{figure} 

Let us first consider the case of an isotropic  s-wave gap singularity. The high frequency modes are still critical, and therefore the high frequency limit of the imaginary part of the pair susceptibility is determined by,
\begin{equation}
\lim_{\omega\to\infty}\chi''(\omega,\Delta,T=0)= \frac{\cal A}{\omega^{\alpha_p}}.
\end{equation}
In the presence of the superconducting condensate, the low energy modes below the gap have their energy raised above the gap, since we require $\chi''(\omega<\Delta,\Delta,T=0)=0$. The spectral weight is conserved according to Eq. (\ref{norm}), and since we assumed that the Bogoliubov excitations  of the BCS fixed point survive at energies of order of the gap we need to incorporate a BCS s-wave type power law divergence right above the gap  in 
the imaginary part of the pair susceptibility. The simplest function satisfying these conditions is,
\begin{equation}
\chi''(\omega,\Delta)={\cal A}\frac{1}{\omega^{\alpha_p}} \left( \frac{\omega}{\sqrt{\omega^2-(2\Delta^2)}} \right)^{1+\alpha_p}\Theta(\omega-2\Delta),
\end{equation}
with ${\cal A}=(1-\alpha_p)\omega_c^{-(1-\alpha_p)}$ (see Fig.5b).  
 We notice in passing that the BCS gap corresponds to the case $\alpha_p=0$,
\begin{equation}
\chi_{\rm BCS}''(\omega,\Delta)=\frac{1}{2E_F} \frac{\omega}{\sqrt{\omega^2-(2\Delta)^2}} \Theta(\omega-2\Delta).
\end{equation}
The quantum critical gap equation for the s-wave superconductor now becomes, 
\begin{equation}
\label{gapm1}
1-2(1-\alpha_p) \lambda\left( \frac{2\omega_B}{\omega_c}\right)^{-\alpha_p}  \left( \frac{\Delta}{\omega_B}\right)^{-\alpha_p}\int_1^{\frac{\omega_B}{\Delta}}\frac{dx}{(x^2-1)^{(1+\alpha_p)/2}}=0.
\end{equation}

Turning to the d-wave case the gap equation becomes necessarily a bit more complicated since we have to account for massless Bogolubiov fermions.  
At low frequencies $\omega<<2 \Delta$ the pair susceptibility is now governed by free fermion loops and the Dirac-cone structure in the spectrum leads to a linear frequency dependence in the pair susceptibility, $\chi''(\omega)={\cal A}_1\omega$. Near the gap, a logarithmic divergence is expected due to the Van Hove singularity, and therefore $\chi''(\omega)={\cal A}_2\log\frac{q_c+\sqrt{2\Delta-\omega+q_c^2}}{-q_c+\sqrt{2\Delta-\omega+q_c^2}}$ for $\omega<2\Delta$, while $\chi''(\omega)={\cal A}_3\log\frac{q_c+\sqrt{\omega-2\Delta+q_c^2}}{-q_c+\sqrt{\omega-2\Delta+q_c^2}}$ for $\omega>2\Delta$, with $q_c$ a cutoff. When the frequency is high compared to the gap scale, the pair susceptibility has the scaling form $\chi''(\omega)={\cal A}_4\omega^{-\alpha_p}$.  Matching these regimes at $2\Delta-\omega_1$ and $2\Delta+\omega_2$, with $0<\omega_1<2\Delta$ and $0<\omega_2<2\omega_B-2\Delta$, and assuming continuity of the pair susceptibility both below and above the gap (see Fig. 5b), we arrive at the gap equation for the d-wave case,
\begin{eqnarray}
\label{gapdd}
\frac{1}{2g}= {\cal A}_1(2\Delta-\omega_1)&+&{\cal A}_2\frac{q_c^2}{2\Delta}\int_0^{\omega_1/q_c^2}\frac{dx}{1-xq_c^2/(2\Delta)}\log\frac{1+\sqrt{x+1}}{-1+\sqrt{x+1}}\nonumber\\     
&+&{\cal A}_3\frac{q_c^2}{2\Delta}\int_0^{\omega_2/q_c^2}\frac{dx}{1+xq_c^2/(2\Delta)}\log\frac{1+\sqrt{x+1}}{-1+\sqrt{x+1}}+\frac{{\cal A}_4}{\alpha_p}\left[(2\Delta+\omega_2)^{-\alpha_p}-(2\omega_B)^{-\alpha_p}\right].
\end{eqnarray}
 This contains a number of free parameters that are partially constrained by the spectral weight conservation. This however does not suffice  to determine the gap uniquely. In the following we will make further choice of the parameters, to plot the gap. We choose  the scaling dimension $\alpha_p=3/4$, and the cut-off in the logarithm to be of order the square root of the gap, say $q_c/\sqrt{2\Delta}=3$, the width of the logarithmic region to be 20 percent of the magnitude of the gap on both sides of the gap, that is $\omega_1/(2\Delta)=\omega_2/(2\Delta)=0.2$,  the coefficient of the high frequency part ${\cal A}_4=1/(4\omega_c^{3/4})$, and further define $\omega_1/q_c^2=\omega_2/q_c^2\equiv a$, 
 $b\equiv\int_0^adx\log\frac{1+\sqrt{x+1}}{-1+\sqrt{x+1}},c\equiv\log\frac{1+\sqrt{a+1}}{-1+\sqrt{a+1}},d\equiv\frac{4\times1.2^{1/4}-1.2^{-3/4}\times9b/c}{0.32+7.2b/c}$,
thus the corresponding d-wave gap equation reads,
\begin{eqnarray}
\label{gapdd1}
1-\frac{1}{2} \lambda\left( \frac{2\omega_B}{\omega_c}\right)^{-\frac{3}{4}}  \left( \frac{\Delta}{\omega_B}\right)^{-\frac{3}{4}}\left(0.8d\right.&+&7.2\frac{d}{c}\int_0^a\frac{dx}{1-9x}\log\frac{1+\sqrt{x+1}}{-1+\sqrt{x+1}}\nonumber\\     
&+&9\frac{1.2^{-\frac{3}{4}}}{c}\int_0^a\frac{dx}{1+9x}\log\frac{1+\sqrt{x+1}}{-1+\sqrt{x+1}}+\frac{4}{3}(1.2^{-\frac{3}{4}}-(\frac{\Delta}{\omega_B})^{\frac{3}{4}})\left.\right)=0,
\end{eqnarray}

We plot in Fig.(6) the behavior of the gap function in the s- and d-wave cases, to be compared with the outcomes Fig.  (2) of the approach taken in section II  where  the gap simply entered as an IR cut-off 
scale, Eq. (\ref{qcgap00}). One can see that in both cases the magnitude of the gap is enhanced by treating the singularity more carefully, while in the d-wave case this enhancement is  even more 
pronounced than in the s-wave case. These effects  can be understood in terms of the redistribution of the spectral weight, since the low frequency part is enhanced by the factor $1/\omega$  
in the Kramers-Kronig frequency integral. The dependence of the gap on the glue strength and retardation does however not change significantly compared to what we found in section II, which can be understood from the fact that the gap depends on the combination $\lambda (2\omega_B/\omega_c)^{-\alpha_p}$. One also notices in Fig.(6) that the magnitude of the gap saturates already at small $\lambda$ for modest retardation. This is an artifact 
of the modeling. In real system the power law (s-wave) or  logarithmic (d-wave)  spectral singularities will be damped (see e.g. \cite{Dzyaloshinskii96,Irkhin01,Irkhin02,Rubtsov09}), and the endpoints at finite $\lambda$ 
in Fig.(6) will turn into smooth functions..

The gap to $T_c$ ratio is expected to be a number order unity number. However, it is quite sensitive to the details of the crossover regime  between the high frequency critical behavior and the low frequency superconducting behavior
as of relevance to the zero temperature gap.  Numerically evaluating Eq.'s (\ref{tc1},\ref{gapm1},\ref{gapdd1}) we obtain gap to $T_c$ ratio's as indicated in Fig. (7). Different from the Migdal-Eliasbergh case we find that 
these ratio's are rather strongly dependent on both the Migdal- and the coupling parameter, while the ratio becomes large for {\em small} coupling, in striking contrast with conventional strong coupling superconductivity.
Invariably we find the ratio to be larger than the weak coupling BCS case, reflecting the strongly dissipative nature of quantum critical states at finite temperature that plays apparently
a similar role as the 'pair-breaking' phonon heat bath in conventional superconductors.

\begin{figure}
\begin{centering}
\includegraphics[width=1.0\linewidth]{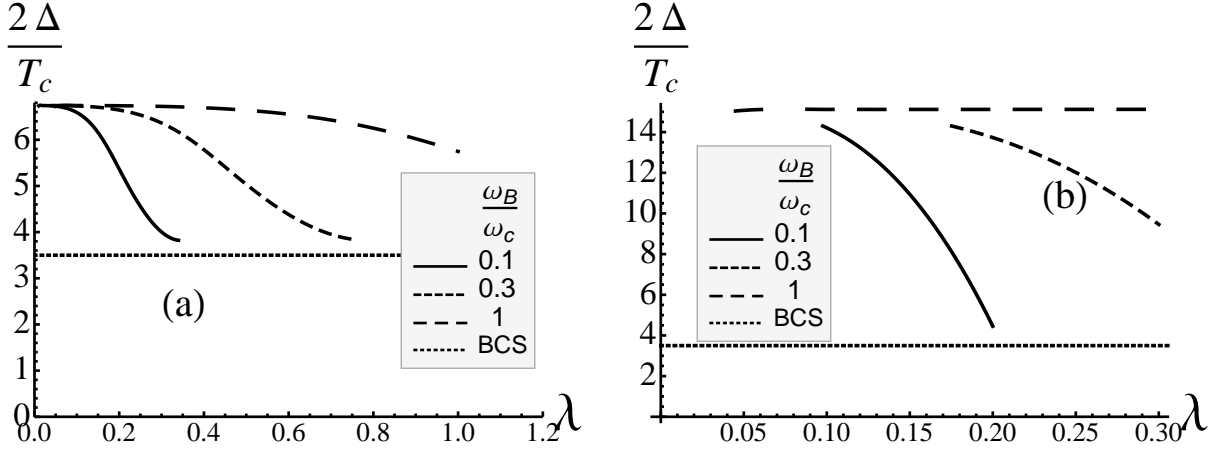}
\end{centering}
\caption{(a)The gap to $T_c$ ratio $2\Delta/T_c$ as a function of glue strength $\lambda$ for  various retardation ranges $\omega_B/\omega_c$ with fixed scaling dimension $\alpha_p=3/4$, for s-wave pairing.  The dotted line is the standard BCS result, where $2\Delta/T_c=3.5$. (b) The same plot for d-wave pairing. The gap to $T_c$ ratio decreases with increasing glue strength and retardation for both s- and d-wave gap. The ratios for different retardation ranges approach the same constant as $\lambda\to 0$.}
\end{figure}

\section{Away from the critical points: the superconducting dome versus $T^*$.}

\begin{figure}
\begin{center}
\includegraphics[width=0.45\linewidth]{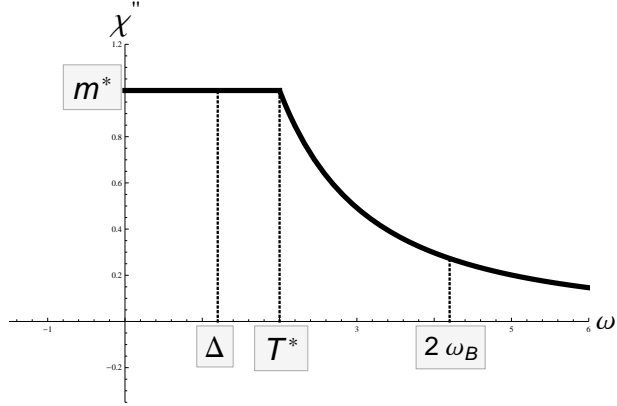}
\end{center}
\caption{Illustration of the imaginary part of the pair susceptibility away from the critical point. For $\omega>T^*$, it has the critical scaling behavior, while for $\omega<T^*$, it retains the BCS form. $T^*$ is the cross-over scale. The effective mass $m^*$ is identified as the magnitude of the imaginary part of the pair susceptibility in the BCS region. The gap $\Delta$ acts as a low energy cut-off, and the retardation $2\omega_B$ as a high energy cut-off. When $T^*$ lies between $\Delta$ and $2\omega_B$, as is the case shown above, both the critical modes and Fermi liquid modes contribute. When $\Delta>T^*$, only the critical modes contribute. When $2\omega_B<T^*$, only the Fermi liquid modes contribute.}
\end{figure}
\begin{figure}
\begin{center}
\includegraphics[width=0.4\linewidth]{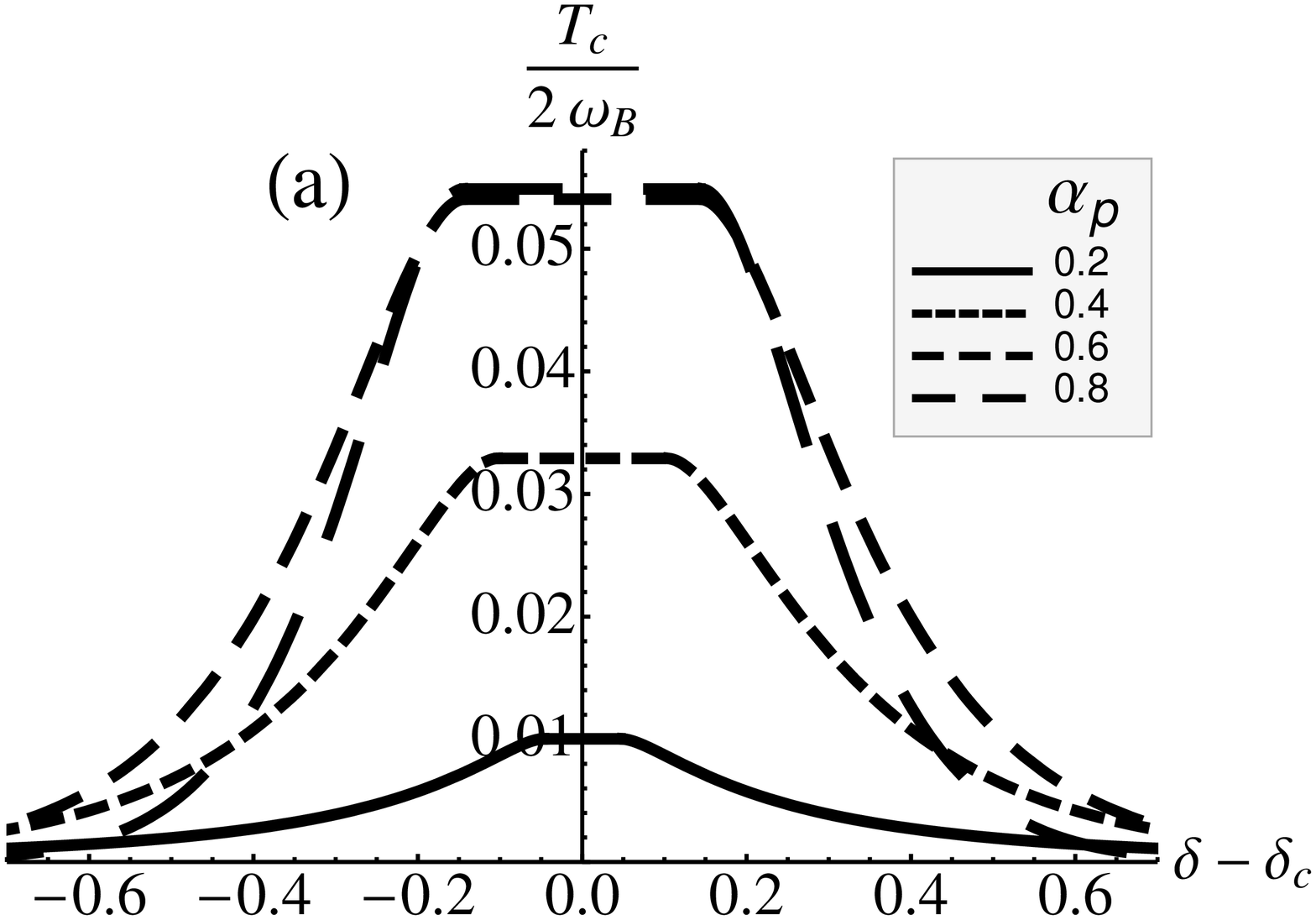}\quad
\includegraphics[width=0.4\linewidth]{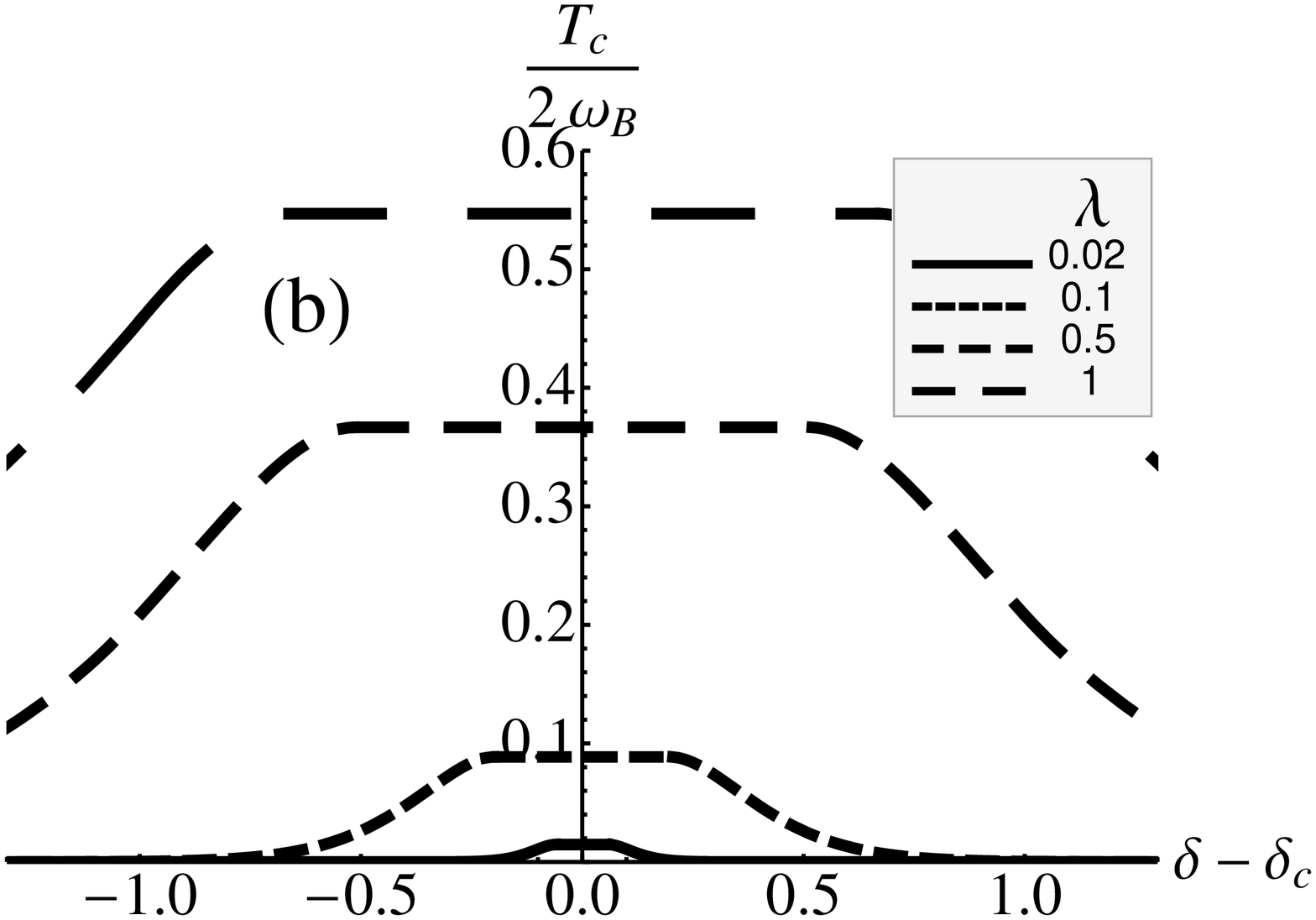}
\includegraphics[width=0.4\linewidth]{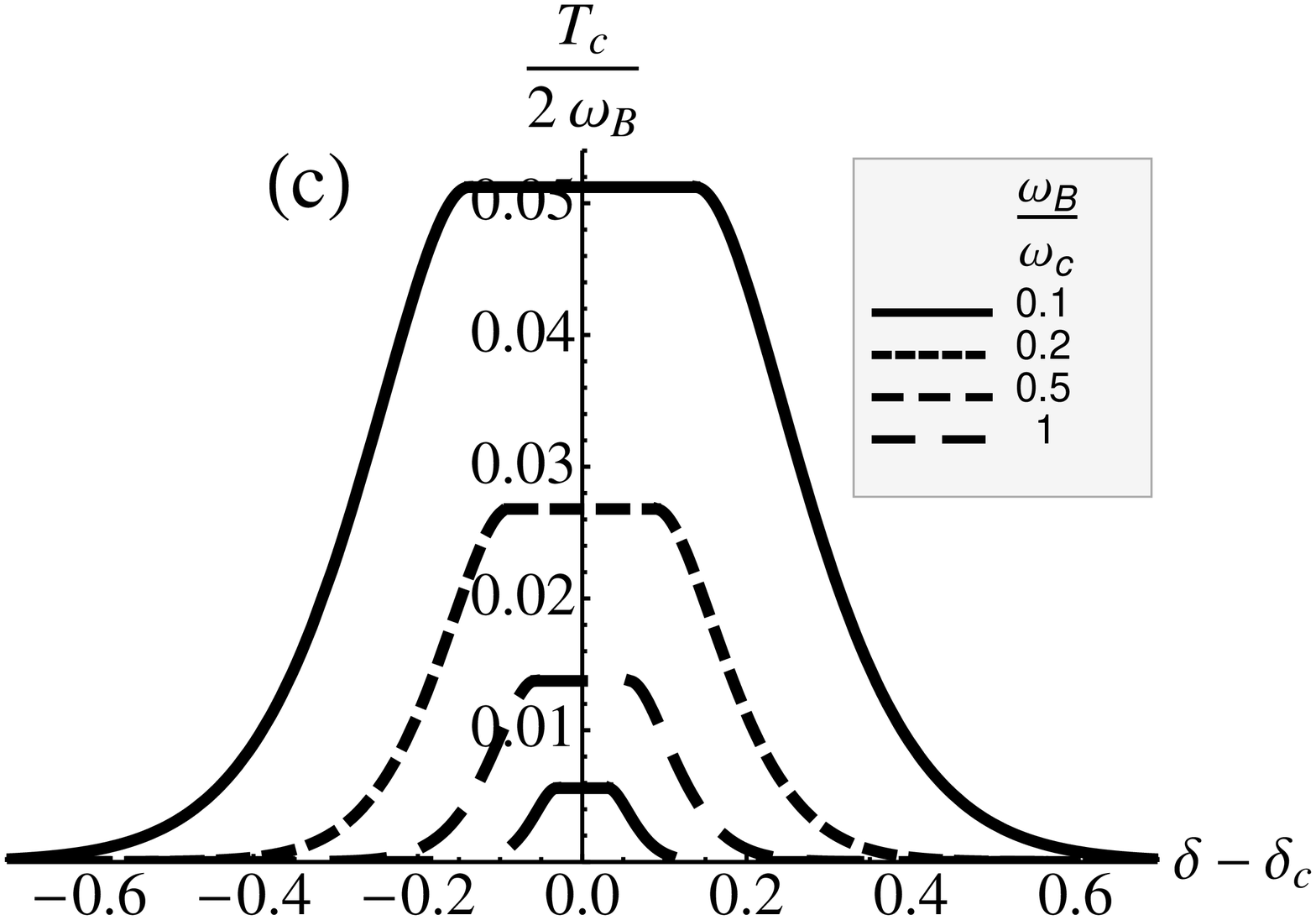}
\includegraphics[width=0.4\linewidth]{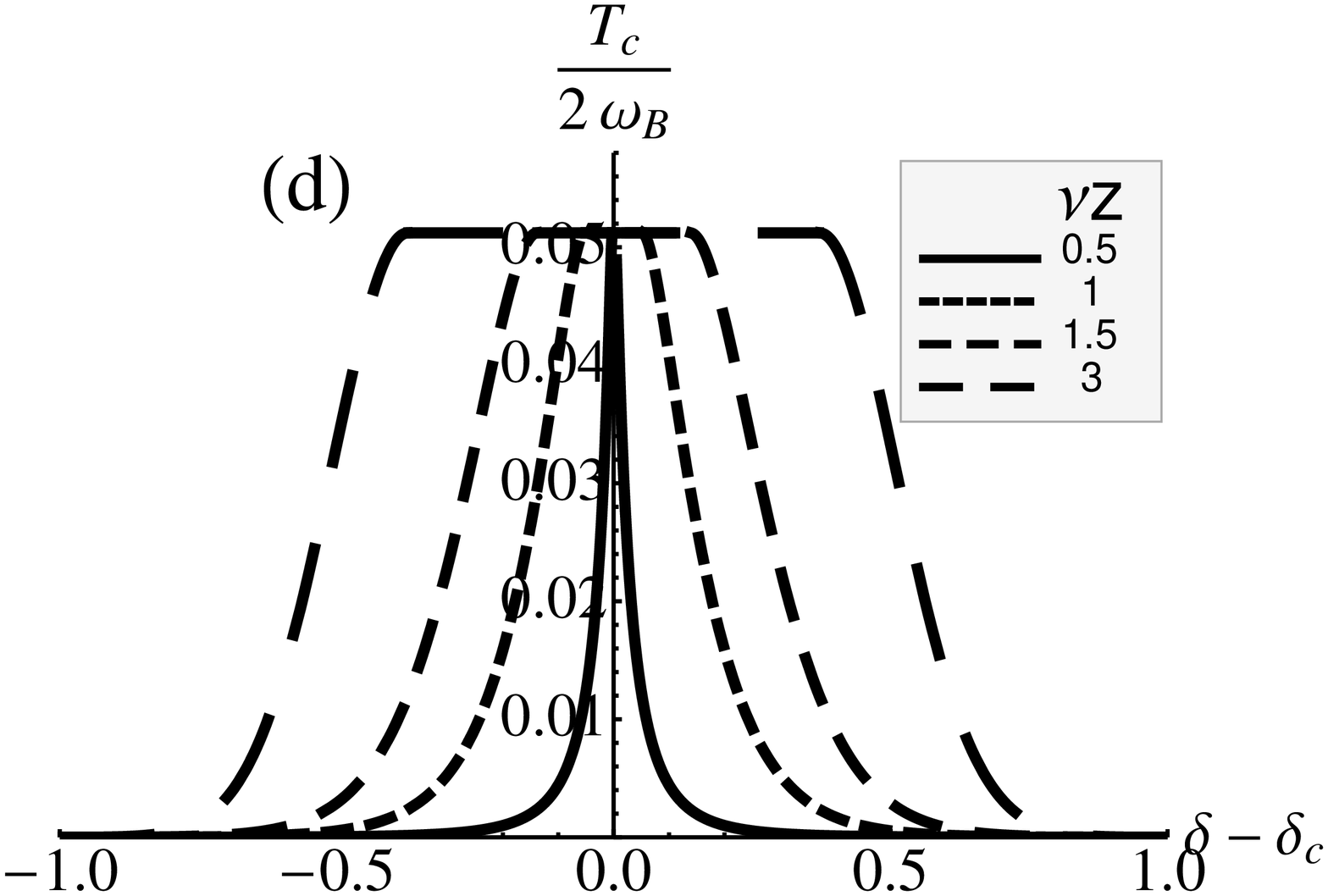}
\end{center}
\caption{The ratio of $T_c$ to retardation as a function of the distance away from criticality (a) for various scaling exponent $\alpha_p$'s with $\lambda=0.06, \omega_B/\omega_c=0.1,\nu z=3/2$, (b) for various glue strength $\lambda$'s with $ \omega_B/\omega_c=0.1,\nu z=3/2,\alpha_p=5/6$. (c) for various retardation over cut-off $\omega_B/\omega_c$'s with $\lambda=0.06, \nu z=3/2, \alpha_p=5/6$. (d) for various inverse Gr\"uneisen exponent $\nu z$'s with $\lambda=0.06, \omega_B/\omega_c=0.1, \alpha_p=5/6$.}
\end{figure}
\begin{figure}
\begin{center}
\includegraphics[width=1.0\linewidth]{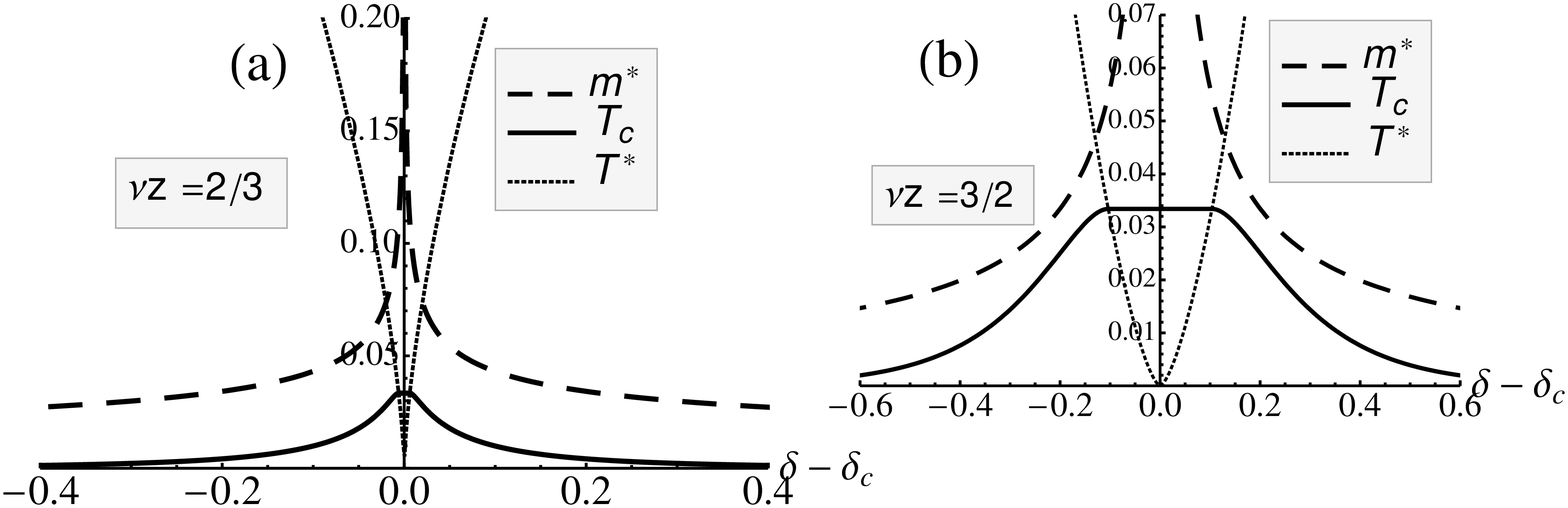}
\end{center}
\caption{(a):The superconducting transition temperature $T_c$ as a function of the distance from the critical point, for given crossover temperature $T^*$ and retardation $\omega_B$. The parameters are chosen as $z=2, \nu=1/3, \eta_p=1,\lambda=0.05, \omega_B/\omega_c=0.1$. (b):The same plot for a different set of parameters $z=3,\eta=0.5, \nu=1/2, \lambda=0.05, \omega_B/\omega_c=0.1$. In-between the two points $\delta_c\pm {\tilde \delta}$, at which the transition temperature coincides with the cross-over temperature $T_c(\delta_c\pm {\tilde \delta})=T^*(\delta_c\pm {\tilde \delta})$, the critical temperature remains constant. For $T^*>2\omega_B$, $T_c$ decays exponentially. The schematic behavior of the effective mass $m^*$ is also included. It diverges when approaching the critical point.}
\end{figure}
Our scaling theory yields a simple and natural explanation for the superconducting domes surrounding the QCP's. This is usually explained
in the Moriya-Herz-Milles framework\cite{Hertz76,Millis93,Moriya95,Pepin05} that asserts that the critical fluctuations of the bosonic order parameter turn into glue with singular strength while the Fermi-liquid
is still in some sense surviving. We instead assert that the glue is some external agent (e.g., the phonons but not necessarily so) that is blind to the critical point, but the 
fermionic criticality boosts the SC instability at the QCP according to Eq. (\ref{qcgap}). By studying in detail the variation of the SC properties in the vicinity of the 
QCP it should be possible to test our hypothesis. The data set that is required is not available in the literature and let us present here a crude sketch
of what can be done.  In at least some heavy fermion systems\cite{Custers03} a rather sudden cross-over is found between the high temperature critical state and
a low temperature heavy Fermi-liquid, at a temperature $T^*\sim |\delta-\delta_c|^{\nu z}$, with $\nu$ 
behaving like a correlation length exponent $\xi\sim|\delta-\delta_c|^{-\nu}$ as function of the zero temperature tuning parameter $\delta$. Moving away from the QPT
this means for the SC instability that an increasingly larger part of the frequency interval of $\chi''$ below $\omega_B$ is governed by the Fermi-liquid 'flow' with the
effect that $T_c$ decreases. We can crudely model this by asserting that
   the imaginary part of the pair susceptibility acquires  the critical form for $\omega>T^*$ and the Fermi-liquid form for $\omega<T^*$, while
we impose that it is continuous at $\omega=T^*$. This model has  the implication that the magnitude of $\chi''$ in the Fermi-liquid regime is determined by $T^*$
and $\eta_p$ and we find explicitly that  $N_0 \propto m^*\propto |\delta-\delta_c|^{-\nu(2-\eta_p)}$. We notice that this should not be taken literally, since this cross-over behavior can be a priori more complicated. In fact, from thermodynamic scaling it is known \cite{Rosch03,Zaanen04} that $m^*\sim  |\delta-\delta_c|^{\nu(d-z)}$. Fig. (8) would imply that $\alpha_p=1-d/z$. This is not implied by scaling.
 
Given these  assumptions, the gap equation away from the quantum critical point becomes, 
\begin{equation}
1-2g\left( \int_{\Delta}^{T^*}\frac{d\omega}{\omega}\chi''_{\rm BCS}(\omega)+\int_{T^*}^{2\omega_B} \frac{d\omega}{\omega}\chi''_{\rm crit}(\omega)  \right)=0
\end{equation}
We are interested in the superconducting transition temperature, which has been shown in the previous section to be approximately the gap magnitude $T_c\simeq \Delta$. 
The imaginary part of the pair susceptibility in the critical region has still the power law form $\chi''_{\rm crit}(\omega)=Z''\sin(\alpha_p \pi/2)\omega^{-\alpha_p}$, while in the BCS region it is a constant determined by continuity at $\omega=T^*$ and therefore $\chi''_{\rm BCS}(\omega)=Z''\sin(\alpha_p \pi/2)(T^*)^{-\alpha_p}$.

Consequently we find  in the regime $T_c<T^*<2\omega_B$ the solution for the gap equation,
\begin{equation}
T_c=2\omega_Bx^{\nu z}\exp\left[ \frac{1}{\alpha_p}\left( 1-x^{\nu(2-\eta_p)}-\frac{1}{\tilde{\lambda}}(\frac{2\omega_B}{\omega_c})^{\alpha_p}x^{\nu(2-\eta_p)} \right) \right],
\label{domeeq}
\end{equation}
where $x^{\nu z}=T^*/(2\omega_B)$.
For $T^*<T_c$ a plateau is found since only the critical modes contribute to the pairing, while for $T^*>2\omega_B$  the BCS exponent takes over since only the (heavy)  Fermi-liquid quasiparticles contribute having as
a consequence,
\begin{equation}
T_c=2\omega_B\exp\left( -(\frac{2\omega_B}{\omega_c})^{\frac{2-\eta_p}{z}}\frac{x^{\nu(2-\eta_p)}}{\alpha_p\tilde{\lambda}}  \right).
\end{equation} 

The outcomes are illustrated in Fig. (9,10).  One notices in all cases that the dome shapes are concave with a tendency for a flat 'maximum'. This is automatically implied by our starting assumptions. When $T_c$ is larger
than $T^*$ only the critical regime is 'felt' by the pairing instability and when this criterium is satisfied   $T_c$ does not vary, explaining the flat maximum. When $T_c$ starts to drop below $T^*$ the superconductivity 
gets gradually depressed because the Fermi-liquid regime increasingly contributes. Eventually, far out in the 'wings', one would still have superconductivity but with transition temperatures that become exponentially small.
The domes reflect just the enhancement of the pairing instability by the critical fermion liquid relative to the Fermi-liquid.

The trends seen in Fig. 9 are easily understood. When the scaling dimension $\alpha_p$ is increasing, i.e. the pair operator is becoming more relevant, the maximum $T_c$ increases while not much happens with 
the width of the dome  (Fig. 9a), for the simple reason that the critical metal becomes more and more unstable towards the superconductor. When the coupling strength $\lambda$ increases  one finds in addition that
the dome gets broader (Fig. 9b)  because the 'contrast' between Fermi-liquid and quantum critical BCS is becoming less, illustrating the surprise that especially weakly coupled quantum critical superconductors are much better 
than their traditional cousins.  The same moral is found back when the Migdal parameter is varied (Fig. 9c), illustrating that at very strong retardation the differences are the greatest. Finally, in Fig. (9d) the evolution 
of the domes are illustrated when one changes the exponents relating $T^*$ to the reduced coupling constant. We find that the dome changes from a quite 'box like' appearance to a 'peak' pending the value of $\nu z$. 
The mechanism  can be deduced from Fig. 10, comparing the situation that the quantum critical 'wedge' is concave (fig. 10a, $\nu z < 1$) with a convex wedge (fig. 10b, $\nu z >1$). Because $T^*$ is varying more slowly
in he latter case with the reduced coupling constant, the quantum critical regime becomes effectively broader with the effect that the quantum critical BCS  keeps control over a wider coupling constant range.  The trends
in Fig.'s (9, 10) are quite generic and it would be interesting to find out whether by systematical experimental effort these behaviors can be falsified or confirmed.

\section{Spatial dependence of the pair susceptibility: upper critical field}

Another experimental observable that should be quite revealing with regard to scaling behavior is 
the orbital limiting upper critical field. The orbital limiting field is set by the
condition that the magnetic length becomes of order of the coherence length, and the latter relates to the 'time like' $T_c$ merely by the dynamical critical exponent $z$. In more detail,
assuming a gap of the form \cite{Rajagopal66},
 \begin{equation}
 \Delta({\vec r})=\Delta_0 \exp\left(-\frac{r^2}{2l^2} \right),
 \end{equation}
  the linearized gap equation in the presence of an orbital limiting magnetic field  becomes \cite{AGD}, 
   \begin{equation}
  \frac{1}{{\Omega^{d-1}g}}=\int_{r_0}^{\infty}K_0(r,\beta)\exp\left( -\frac{r^2}{2l^2}\right)r^{d-1}dr,
 \end{equation}  
where $\Omega^{d-1}$ is the volume of the $d-1$-dimensional unit sphere, $l$ is the magnetic length related to the field by $H={\phi_0}/({2\pi l^2})$ where $\phi_0=hc/e$, while $K_0(r,\beta)$ is the real space pair susceptibility, which is the Fourier transform of $\chi'$\cite{Dias94,Schofield95}. For free fermions, the real space pair susceptibility is (see eg. \cite{Dias94}),
\begin{equation}
K_0(r,\beta)=\left( \frac{k_F}{2\pi r} \right)^{d-1}\frac{1}{v_F^2\beta}\frac{1}{\sinh(\frac{2\pi r}{\beta v_F})},
\end{equation}
with a power law behavior $K_0(r,\beta)\sim r^{-d}$ at short distances or low temperatures where $r<\beta v_F$, and an exponential decay at large distances or high temperature. Let us consider critical fermions at $T=0$, such that  the pair susceptibility has the power law form $\chi(\omega)\sim \omega^{-(2-\eta)/z}$. The momentum dependence can be determined by replacing $\omega$ by $k^z$, such that $\chi(k)\sim k^{-(2-\eta)}$. It follows that the real space pair susceptibility has the power law form $K_0(r,T=0)\sim \int \chi(k) \exp(i{\vec k}\cdot{\vec r})d^d{\vec k}\sim r^{-(d-2+\eta_p)}$.
Associate with the retardation a short distance cutoff $r_0$, and assume a scaling $2\omega_B/\omega_c=(r_0/a_c)^{-z}$, where $a_c$ is the lattice constant. The magnetic length acts as a long distance cutoff and therefore,
\begin{equation}
  \frac{1}{{\Omega^{d-1}g}}=\int_{r_0}^{l}\frac{{\cal C}_h}{r^{d-2+\eta_p}}r^{d-1}dr,
\end{equation}
with the normalization factor ${\cal C}_h\simeq  2z(1-\alpha_p)\Omega^{-(d-1)}{\omega_c}^{-1}a_c^{-(2-\eta)}$, so that $(1/\Omega^{d-1})\int_{a_c}K_{\rm crit}(r)r^{d-1}dr\simeq \frac{1}{\omega_c}$, to give the right scale.
The zero temperature upper critical field has then the same form as the one for $T_c$ except for the occurrence of $z$,
\begin{equation}
\label{qchc2}
\frac{2\pi H_{c2}(0)}{\phi_0 r_0^{-2}}\simeq  \left({1+\frac{1}{\tilde{\lambda}}\left(\frac{2\omega_B}{\omega_c} \right)^{\alpha_p} } \right) ^{-\frac{2}{2-\eta_p}},
\end{equation}
 and it follows,
\begin{equation}
\frac{2\pi H_{c2}(0)}{\phi_0 a_c^{-2}}\simeq \left(\frac{ T_c}{\omega_c}\right) ^{2/z}.
\end{equation}
In the BCS case one has $H_{c2}(0)/({\cal B}\phi_0 k_F^{2})= ( T_c/E_F)^2$,
with ${\cal B}\simeq3.26$ for $d=3$ \cite{Helfand66}.  The moral is obvious: in 
Lorentz-invariant ($z=1$) systems the relation between $H_{c2}$ and $T_c$ is the same as for standard BCS, but when the normal
state is governed by a universality class characterized by $z >1$, $H_{c2}(0)$ will be amplified for a given $T_c$ relative to 
conventional superconductors because $T_c/\omega_c,T_c/E_F\ll 1$. 

\begin{figure}
\begin{center}
\includegraphics[width=0.4\linewidth]{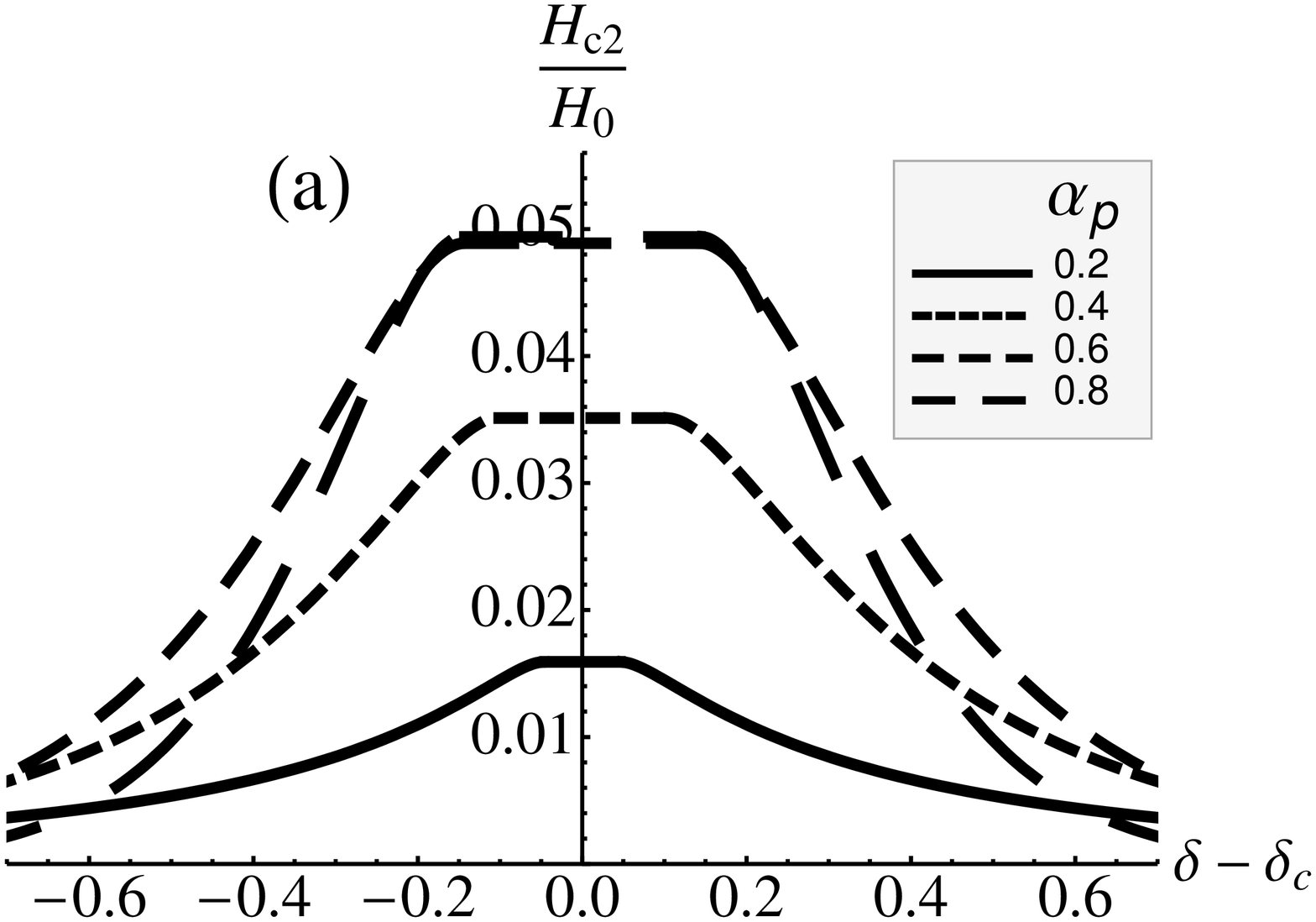}\quad
\includegraphics[width=0.4\linewidth]{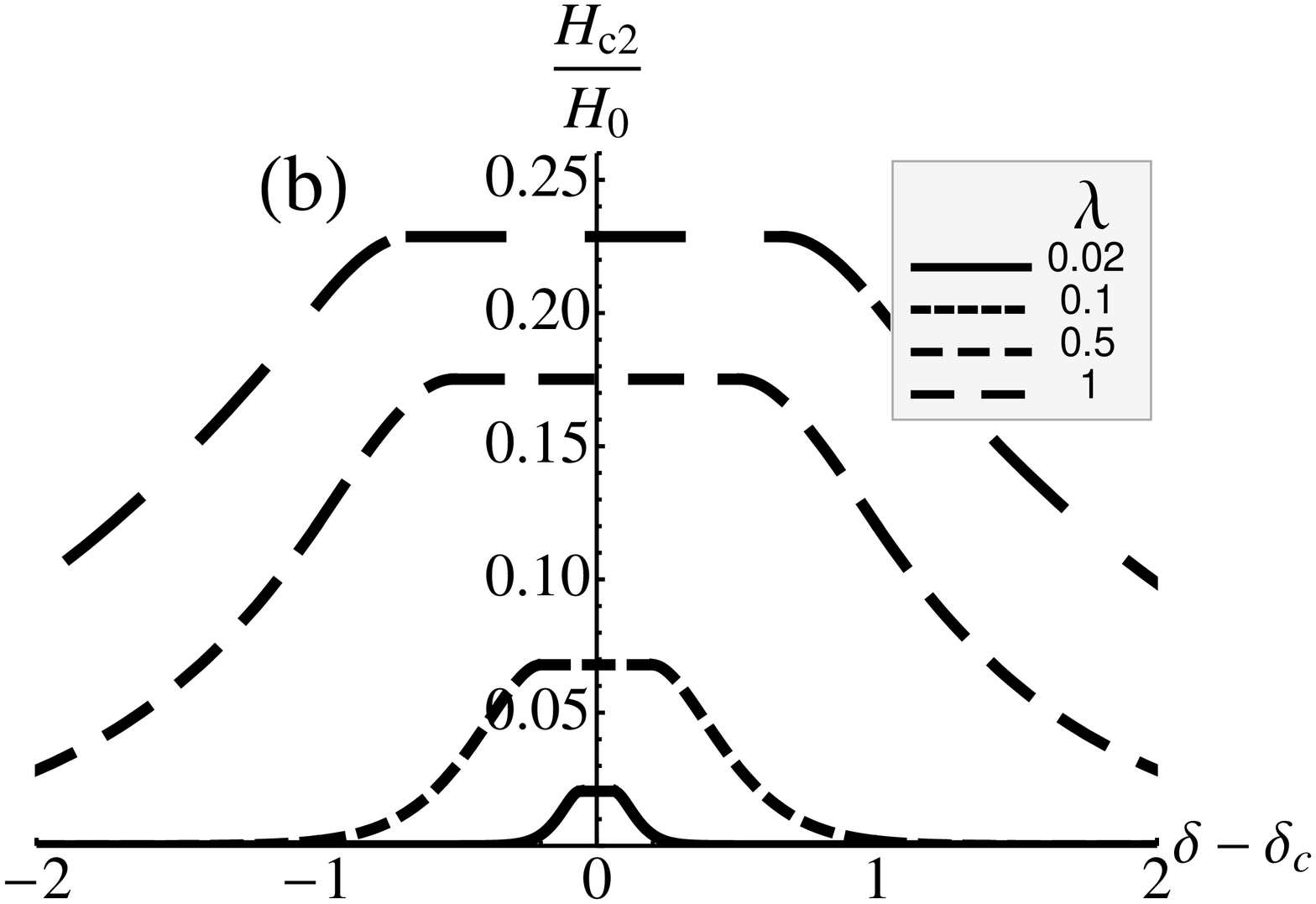}
\includegraphics[width=0.4\linewidth]{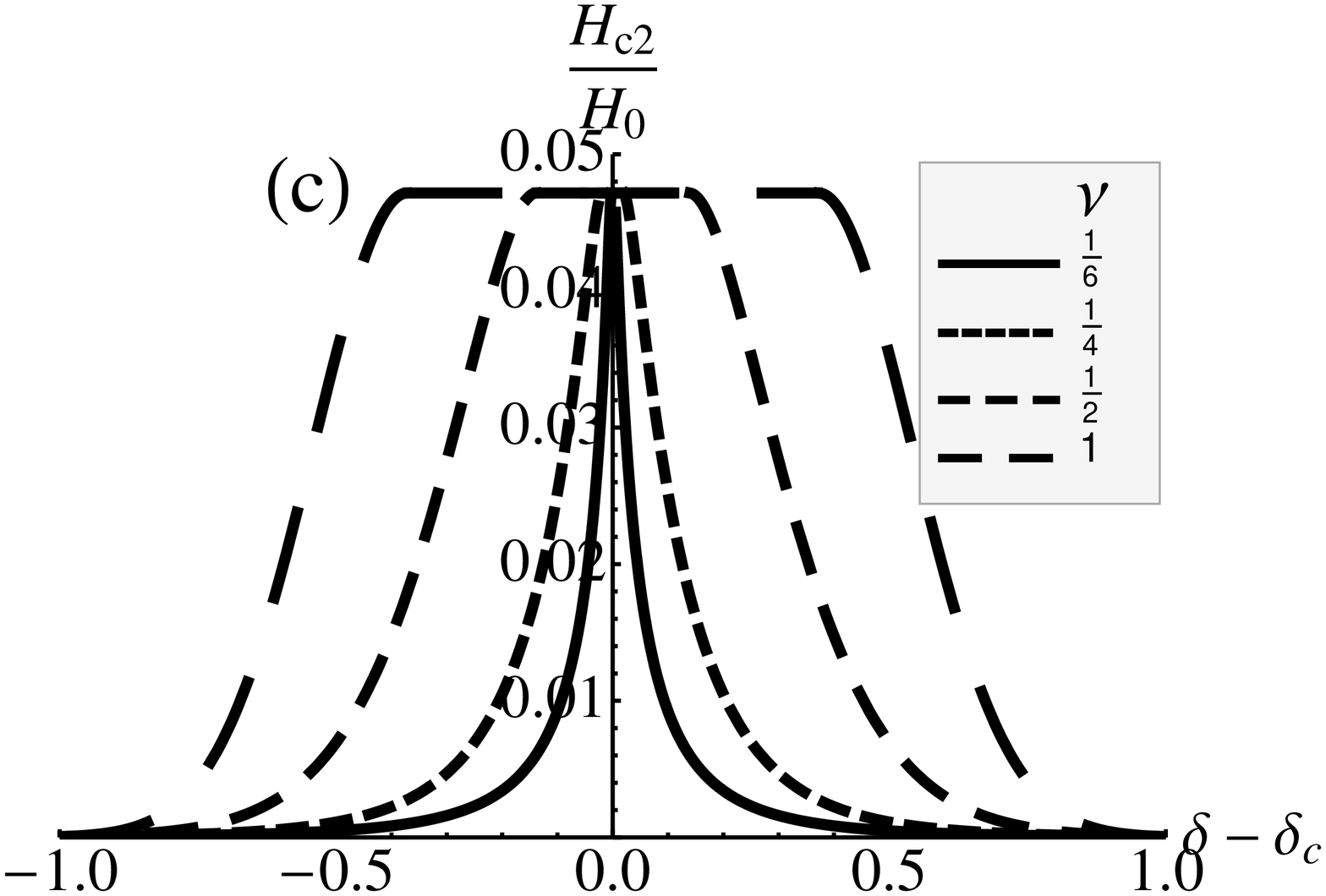}
\includegraphics[width=0.4\linewidth]{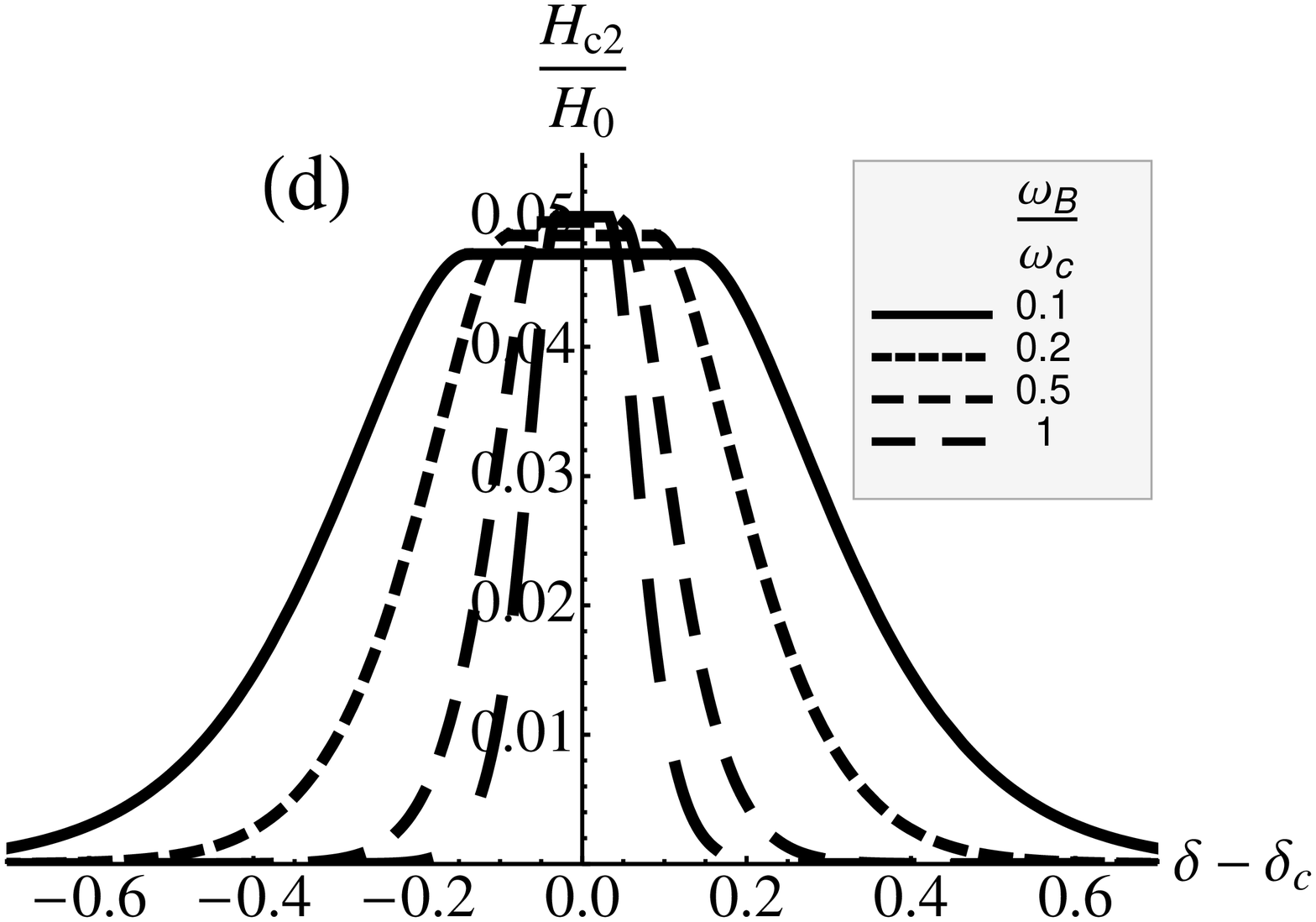}
\end{center}
\caption{The upper critical field $H_{c2}$ over $H_0\equiv \phi_0 a_c^{-2}/(2\pi )$ as a function of the distance away from criticality (a) for various scaling exponent $\alpha_p$'s with $\lambda=0.06, \omega_B/\omega_c=0.1,\nu =1/2,z=3$, (b) for various glue strength $\lambda$'s with $ \omega_B/\omega_c=0.1,\nu =1/2,z=3,\alpha_p=5/6$, (c) for various $\nu$'s with $\lambda=0.06, \omega_B/\omega_c=0.1, \alpha_p=5/6,z=3$, (d) for various retardation ranges with $\lambda=0.06, \nu=1/2,z=3, \alpha_p=5/6$.}
\end{figure}
\begin{figure}
\begin{center}
\includegraphics[width=0.45\linewidth]{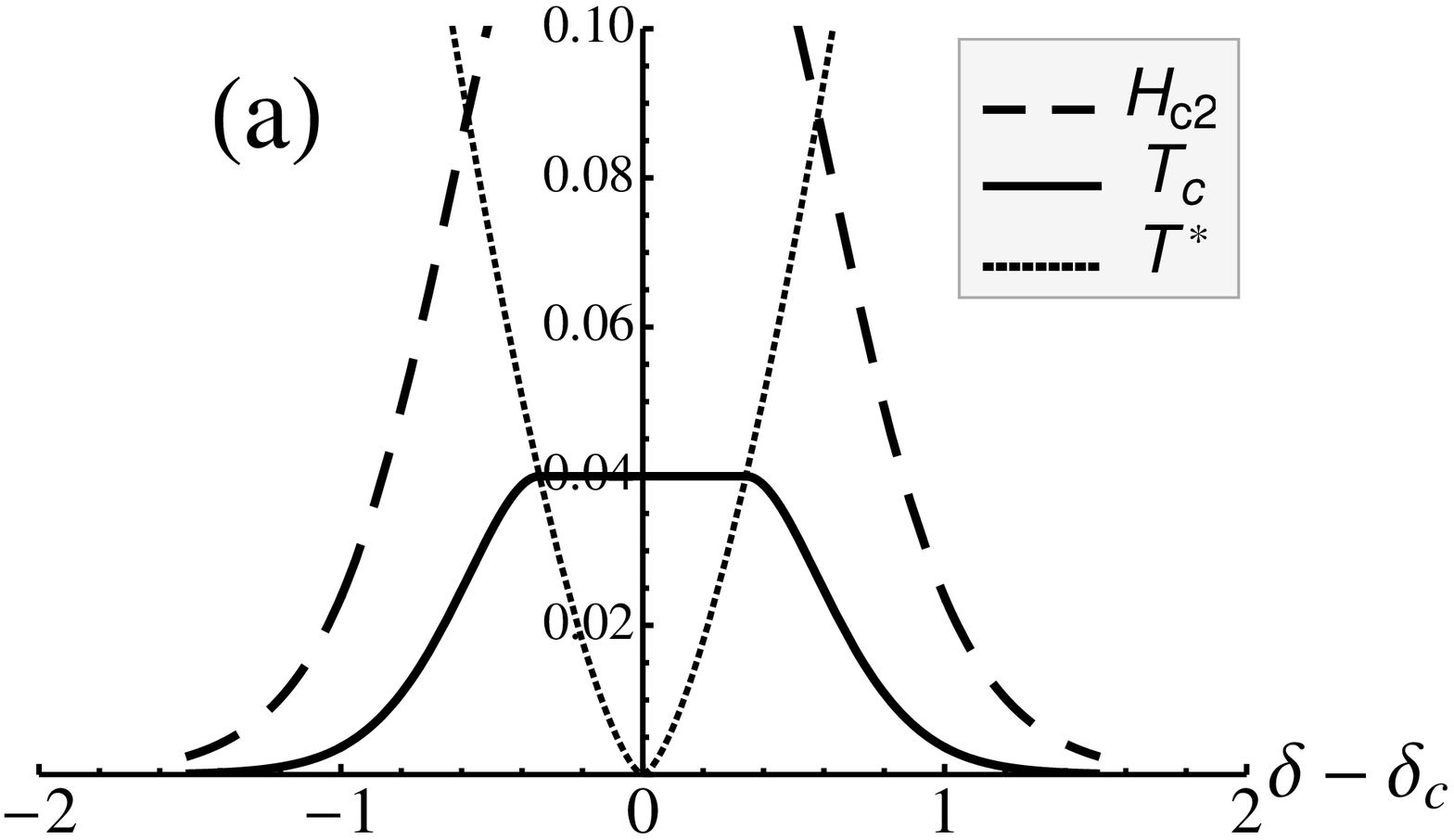}\quad
\includegraphics[width=0.45\linewidth]{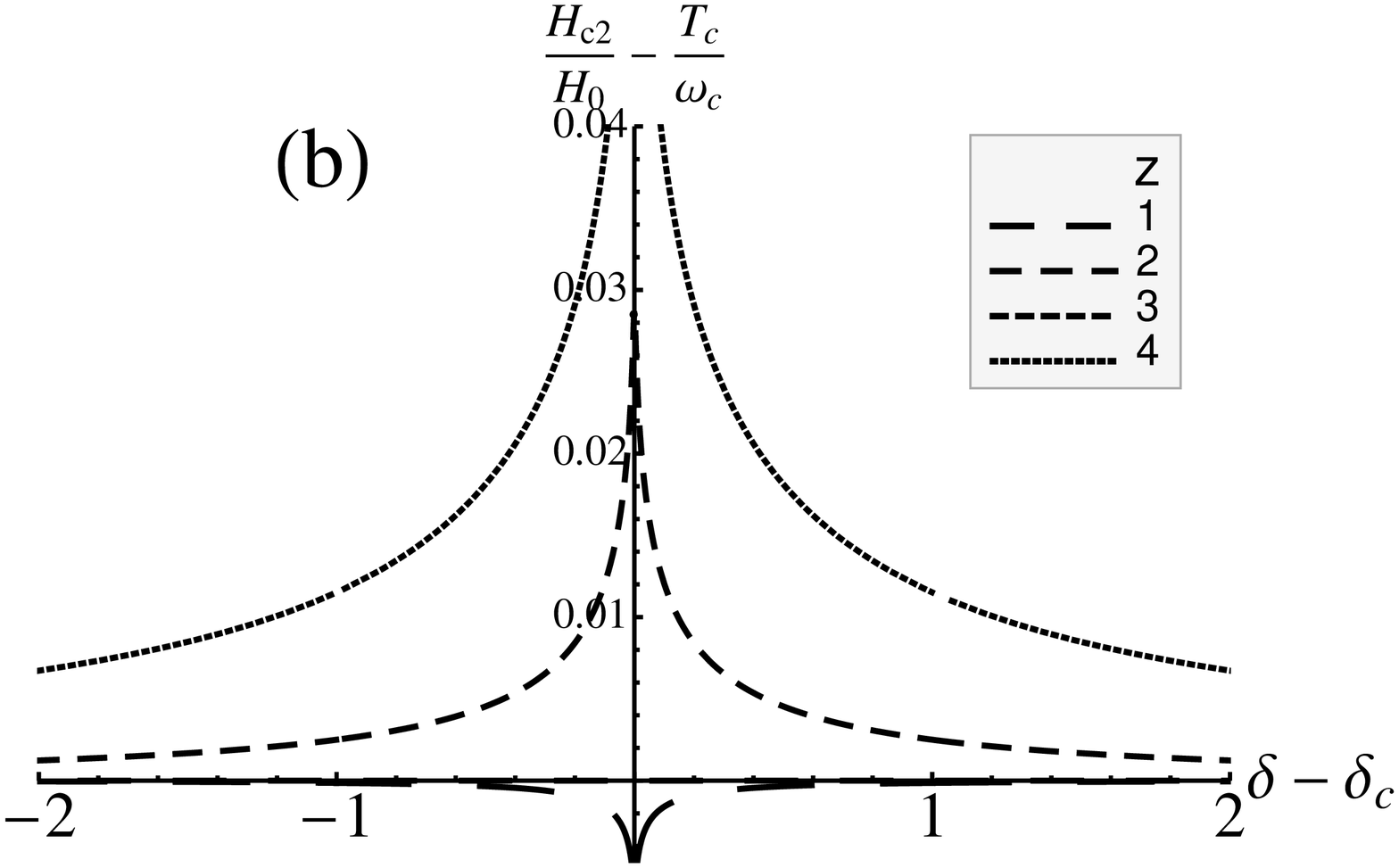}
\end{center}
\caption{(a)Illustration of the different behavior of $T_c$ and upper critical field $H_{c2}$ as the quantum critical point is approached. $H_{c2}$ increases much faster than $T_c$. Thus for a small $T_c$ one can still have a large upper critical field. Here we plotted using the parameters $\lambda=0.05, \omega_B/\omega_c=0.1,\nu =1/2,z=3,\eta=-1$. (b) The difference $H_{c2}/H_0-T_c/\omega_c$ as a function of the distance away from the critical point for different dynamical exponent $z$'s. Here $H_0\equiv \phi_0 a_c^{-2}/(2\pi )$, $\lambda=0.06, \omega_B/\omega_c=0.1,\nu z=0.5,\alpha_p=0.4$. For $z=2$, the difference is 0. For $z=3,4$, the difference is positive and increases rapidly when approaching the critical point. For the case with $z=1$, the difference is negative.}
\end{figure}

Modeling the variation of $H_{c2}$ in the vicinity of the QPT as in the previous paragraph, where the critical modes govern the short distance and BCS type behavior is recovered at large distance, while converting the cross-over temperature to a length scale $r^*$, by $T^*/\omega_c=(r^*/a_c)^{-z}$, 
we find that $H_{c2}$ is determined by the equation,
\begin{equation}
  \frac{1}{{\Omega^{d-1}g}}=\int_{r_0}^{r^*}\frac{{\cal C}_h}{r^{d-2+\eta_p}}r^{d-1}dr+\int^{l}_{r^*}\frac{{\cal C}'_h}{r^{d}}r^{d-1}dr,
\end{equation}
with the matching condition ${\cal C}_h=(r^*)^{-2+\eta_p}{\cal C}'_h$.
We find that 
one just has to replace the first two dynamic exponent $z$'s in Eq. (\ref{domeeq}) by $2$ while an extra factor of 2 has to be added to the second term in the exponent, 
\begin{equation}
H_{c2}=\frac{\phi_0 a_c^{-2}}{2\pi }x^{2\nu } \left( \frac{2\omega_B}{\omega_c}\right)^{2/z}   \exp\left[ \frac{2}{2-\eta_p}\left(1-x^{\nu(2-\eta_p)}- (\frac{2\omega_B}{\omega_c})^{\alpha_p}\frac{x^{\nu(2-\eta_p)}}{\tilde{\lambda}}\right) \right].
\end{equation}
In the region where only the Fermi-liquid quasiparticles contribute,
the upper critical field has still an exponential form, \begin{equation}
H_{c2}=\frac{\phi_0 a_c^{-2}}{2\pi } \left( \frac{2\omega_B}{\omega_c}\right)^{2/z}   
\exp\left[ - 2(\frac{2\omega_B}{\omega_c})^{\frac{2-\eta_p}{z}}\frac{x^{\nu(2-\eta_p)}}{(2-\eta)\tilde{\lambda}} \right].
\end{equation}
The dependence of $H_{c2}$ on various parameters is shown in Fig. 11, and one infers that $H_{c2}$ behaves in ways very similar $T_c$ (Fig. 10). The interesting part is illustrated  in Fig.(12b) where we plot $H_{c2}/H_0-T_c/\omega_c$ as a function of the distance away from the critical point for different dynamical exponent $z$'s, keeping all other quantities fixed, defining $H_0\equiv \phi_0 a_c^{-2}/(2\pi )$. One infers that when $z>2$, $H_{c2}/H_0-T_c/\omega_c$ increases rapidly when approaching the critical point. 

Using a 'ferromagnetic' 
dynamical exponent $z=3$  and a  Gr\"uneisen exponent $1/\nu z=2/3$ inspired on recent experiments \cite{Si03,Tokiwa09} as well as theoretical considerations \cite{Hertz76,Millis93,Moriya95,Si01,Pepin05,Senthil08}
we obtain the results in Fig. (12a). Compared to $T_c$, $H_{c2}$ peaks much more strongly towards the QCP. This is in remarkable qualitative agreement with the recent results by Levy {\em et al.} on the behavior of the
orbital limiting field in ${\rm URhGe}$ exhibiting a ferromagnetic QCP\cite{Levy07}, where the highest $T_c$ is about 0.5 K \cite{Levy05}, while the upper critical field exceeds 28 T. It has also been observed in noncentrosymmetric heavy fermion superconductors ${\rm CeRhSi}_3$ \cite{Kimura05, Muro07} and ${\rm CeIrSi}_3$ \cite{Sugitani06,Okuda07}, where the Pauli limiting effect is suppressed due to lack of inversion center of the crystal structures and the orbital limiting effect plays the main role of pair breaking. Near the quantum critical points, $H_{c2}$ can be as high as about 30 K, although the zero field $T_c$ is of order  1K \cite{Kimura07,Settai08}. This class of experiments can be understood in our framework as resulting from the change of the scaling relation between $H_{c2}$ and $T_c$. (See also \cite{Tada08} for a tentative explanation from the customary Hertz-Millis-Moriya perspective.)

\section{Conclusions}

Perhaps the real significance of the above arguments is no more than to supply a cartoon, a metaphor to train the minds on thinking about pairing 
instabilities in non Fermi-liquids. This scaling theory has the merit of being mathematically controlled, given the starting assumptions of the 'retarded
glue' and conformal invariance. The Migdal parameter plays an identical role as in conventional BCS theory to  yield a full control over the glue-fermion
system dynamics, while we trade in Fermi-liquid principle for the even greater powers of scale invariance. The outcomes are gap and Tc equations where 
the standard BCS/Eliasberg equations show up as quite special cases associated with the marginality of the pair operators of the Fermi gas. The difficulty 
is of course to demonstrate that these starting assumptions have dealings with either nature itself and/or microscopic theories of electron systems 
where they should show up as emergent phenomena at low energy. However, the same objections apply to much of the current thinking regarding superconducting
instabilities at quantum critical points with their implicit referral to a hidden Fermi gas. In such considerations there is an automatism to assume that eventually
the superconductivity has to be governed by Eliashberg type equations. At the least, the present analysis indicates that such equations are not divine as long as the
Fermi-liquid is not detected directly. Stronger, in line with the present analysis one might wish to conclude that superconducting instabilities will
be generically more muscular in any non-Fermi-liquid. The Fermi-liquid is singular in the regard that its degrees of freedom are stored in the Fermi-sea, and this
basic physics is responsible for the exponential smallness of the gap in terms of the coupling constant. This exponential smallness should be alien to any non Fermi-liquid.

How about experiment? Scaling theories have a special status in physics because they guide the analysis of experimental data in terms of a minimal a-priori 
knowledge other than scale invariance. The present theory has potentially the capacity to produce high quality empirical tests in the form of scaling collapses.
However, there is a great inconvenience: one has to be able to vary the glue coupling strength, retardation parameters and so forth, at will to test the scaling 
structure of the equations. These are parameters associated with the materials themselves, and one runs into the standard difficulty that it is impossible to vary these in
a controlled manner.  What remain are the rather indirect strategies discussed in the last two sections: find out whether hidden relations exist between the
detailed shape of the superconducting and the crossover lines; are there scaling relations between $H_{c2}$ and $T_c$ as discussed in the last section? 
We look forward to experimental groups taking up this challenge.  

There appears to be one way to interrogate our starting assumptions in a very direct way by experiment. Inspired by theoretical work by Ferrell \cite{Ferrell69} and Scalapino \cite{Scalapino70}, Anderson and
Goldman showed quite some time ago \cite{Goldman70} that the dynamical pair susceptibility can be measured directly using the AC Josephon effect -- see also \cite{Takayama71,Yoshihiro70}, for a recent review see ref.\cite{Goldman06}.
It would be interesting to find out whether this technique can be improved to measure the pair susceptibility over the large frequency range, 'high' temperatures and
high resolution to find out whether it has the conformal shape. It appears to us that the quantum critical heavy fermion superconductors offer in this regard better
opportunities than e.g. the cuprates given their intrinsically much smaller energy scales.

In conclusion, exploiting the motives of retardation and conformal invariance we have devised a phenomenological scaling theory for superconductivity
that generalizes the usual BCS theory to non Fermi-liquid quantum critical metals. The most important message of this simple construction is that it demonstrates
the limitations of the usual Fermi-liquid BCS theory. The exponential smallness of the gap in the coupling is just reflecting the 'asymptotic freedom' of the Fermi-liquid,
and this is of course a very special case within the landscape of scaling behaviors. Considering the case that the pair operator is  relevant, we find instead an 'algebraic'
gap equation revealing that at weak couplings and strong retardation the rules change drastically: as long as the electronic UV cut-off and the glue energy are large, one
can expect high $T_c$'s already for quite weak electron-phonon like couplings. If our hypothesis turns out to be correct, this solves the problem of superconductivity at a high temperature although 
it remains to be explained why quantum critical normal states can form with the required properties. It is however not straightforward to device a critical test for our hypothesis.
The problem is the usual one that pair susceptibilities, $\lambda$'s or $\alpha^2 F$'s, and so forth cannot be measured directly and one has to rely on imprecise modelling.
However, it appears to us that 'quantum critical BCS superconductivity' works so differently from the Fermi-liquid case that it eventually should be possible to nail it down 
in the laboratory. We hope that the sketches in the above will form a source of inspiration for future work.

{\em Acknowledgements} We acknowledge useful discussions with M. Sigrist, D. van der Marel, A.V. Balatsky, B. J. Overbosch, D.J. Scalapino, K. Schalm and S. Sachdev.    This work is supported by the 
Nederlandse Organisatie voor Wetenschappelijk Onderzoek (NWO) via a Spinoza grant.

\bibliographystyle{apsrev}
\bibliography{strings,refs}

\end{document}